\documentclass[preprint2]{aastex}
\usepackage{mkfig}
\accepted{}

\newcommand{\erf}{\mbox{erf}}
\shorttitle{Mixing in the ISM}
\shortauthors{Avillez \& Mac Low}

\received{2001 December 14}
\begin{document}

\title{Mixing Time Scales in a Supernova-Driven Interstellar Medium}

\author{Miguel A. de Avillez \& Mordecai-Mark Mac Low}
\affil{Department of Astrophysics, American Museum of Natural History, \\
    Central Park West at 79th Street, New York, NY 10024, U.S.A.}
\email{mavillez@amnh.org, mordecai@amnh.org}
\begin{abstract}
We study the mixing of chemical species in the interstellar medium
(ISM).  Recent observations suggest that the distribution of species
such as deuterium in the ISM may be far from homogeneous. This raises
the question of how long it takes for inhomogeneities to be erased in
the ISM, and how this depends on the length scale of the
inhomogeneities. We added a tracer field to the three-dimensional,
supernova-driven ISM model of Avillez (2000) to study mixing and
dispersal in kiloparsec-scale simulations of the ISM with different
supernova (SN) rates and different inhomogeneity length scales.  We
find several surprising results.  Classical mixing length theory fails
to predict the very weak dependence of mixing time on length scale
that we find on scales of 25--500~pc.  Derived diffusion coefficients
increase exponentially with time, rather than remaining constant.  The
variance of composition declines exponentially, with a time constant
of tens of Myr, so that large differences fade faster than small ones.
The time constant depends on the inverse square root of the supernova
rate.  One major reason for these results is that even with numerical
diffusion exceeding physical values, gas does not mix quickly between
hot and cold regions.

\end{abstract}


\section{Introduction}

Measurements of the O/H ratio in the interstellar medium (ISM) along lines 
of sight reaching altitudes above the galactic disk of 
up to 1500 pc by Meyer et
al.\ (1998) and Cartledge et al.\ (2001) suggest that the ISM is
well-mixed up to these heights. This suggestion is also supported by
the O/H ratios observed in 41 H{\rm II} regions in M101 (Kennicutt \&
Garnett, 1996), which show only a 0.1--0.2 dex dispersion about the
radial gradient. In NGC 1569 and 4214 the dispersion is extremely low
(Kobulnicky \& Skillman 1996, 1997). Based on these observations it
seems that there exists strong evidence pointing to a fairly
well-mixed ISM in galaxies including the Milky Way. 

On the other hand, recent observations using the 
Interstellar Medium Absorption Profile
Spectrograph (IMAPS) suggest that there are local variations in the D/H
ratio of up to a factor of three among the studied lines of sight
(Jenkins et al.\ 1999; Sonneborn et al.\ 2000; for an up to date
discussion on deuterium abundances see York 2002). The detection of
deuterium in the ISM indicates that it was not processed into
stars and the local variations reported in the D/H ratio suggest
incomplete mixing in the ISM. However, such an inhomogeneous picture
contradicts the almost homogeneous distribution seen in O/H ratioes.

Deuterium was produced during big bang nucleosynthesis and is destroyed
during the pre-main sequence phase of stars. Thus, its ratio to
hydrogen is initially constant in primordial gas 
and decreases as it is processed in
stars. Deuterium may also be added into galactic interstellar gas 
by accretion of
primordial gas, as required by some chemical evolution models in order
to explain its present abundance in the Galaxy (see Tosi
et al.\ 1998 for details.).
Mulan \& Linsky (1999) proposed that deuterium could be formed
in stellar flares from M dwarf stars. However, the production of
deuterium by this mechanism should be accompanied 
by a significant production of
lithium, which has not been observed.

Contrary to oxygen, deuterium has a primordial origin and limited
destruction mechanisms, making this species a good tracer of the
mixing in the ISM. Abundances of deuterium are probably not influenced
by depletions onto grains, as is the case with most other
species. Although there has been some suggestion that deuterium may be
preferentially bonded onto the surfaces of dust grains, this effect is
not sufficient to explain the large variations in D/H seen in the
ISM. Furthermore, deuterium and oxygen are the only species that have
an ionization fraction that closely follows that of H. Deuterium and
hydrogen have almost the same ionization potentials, photoionization
cross sections and recombination coefficients, making it difficult for
the D/H variations to be a result of any unusual ionization
effects. Other effects such as fractionation by molecular chemistry
have been considered by different authors, but there is no strong
evidence that such effects can be responsible for the D/H variations
seen in the ISM.

We face a set of observations that point towards contradicting
conclusions. O/H observations support the idea that 
complete mixing occurs in the
ISM on timescales comparable to those for chemistry to reach
equilibrium, while D/H measurements indicate a not so well mixed
ISM. Solving this contradiction is an important step towards the
understanding of the mixing and variation of these elements in the
ISM, and therefore, of the the interpretation of local measurements as
constraints of its cosmic abundance.

Roy \& Kunth (1995) discussed the problem of mixing in the ISM and
concluded that the ISM should appear more homogeneous than it does.
These authors suggested that mixing in the ISM is driven by three main
mechanisms acting on different length scales. On the Galactic scale,
they argued that 
turbulent diffusion of interstellar clouds in the shear flow of
galactic differential rotation could remove the inhomogeneities
in the ISM in less than $10^{9}$ yr, in particular the fluctuations
seen in O/H. At scales of $100\le \ell \le 1000$ pc, cloud collisions
and expanding supershells driven by supernovas (SNs) from evolving massive star
associations, differential rotation and triggered star formation (by
stellar winds and SN explosions) could re-distribute and mix inhomogeneities
efficiently in about $10^{8}$~yr; and at smaller scales of the order of $1
\le \ell\le 100$ pc, turbulent diffusion may be the dominant mechanism
in cold clouds, while Rayleigh-Taylor and Kelvin-Helmoltz
instabilities could develop in regions of gas ionized by massive stars,
leading to complete mixing in less than 2~Myr. (This analysis does not
take into account that during the mixing process new inhomogeneities
are added into the ISM.)

Bateman \& Larson (1993) already pointed out that turbulent mixing
should be important in the ISM. However, they based their analysis on
a picture of cloud collisions that ignores the likely origin of clouds
as density enhancements in regions of converging flow in the turbulent
ISM (e.g. V\'azquez-Semadeni et al.\ 1995).

Tenorio-Tagle (1996) noted that classical diffusion theory on the
molecular level predicts mixing timescales in the ISM for gas with
$T<10^{5}$~K several orders of magnitude longer than that
proposed by Roy \& Kunth (1995). As a consequence, Tenorio-Tagle (1996)
concluded that turbulent mixing as well as hydrodynamical
instabilities, as proposed by Roy \& Kunth (1995), are important mixing
mechanisms that allow the reduction of the long mixing time scales
obtained from classical diffusion theory.

The observations showing that the ISM has an inhomogeneous
distribution pose the question of how it is possible in a SN-driven
ISM for different species not to be mixed well enough to suppress local
variations of their ratio to hydrogen along and among lines of
sight. The ISM is regulated by SN explosions, and therefore, by
well-structured and explosive flows rather than just by diffuse
turbulence that acts on the smaller scales. To understand the mixing
process in such a medium we carry out direct numerical simulations of
the evolution of the ISM polluted with inhomogeneities of different
length scales.  We aim to understand if classical mixing length theory
can be used to predict the mixing time scales as a function of size scale.

In \S~2 we review the existing theoretical ideas on dispersion,
diffusion and turbulent mixing and apply them to the ISM. Section 3
describes the modified version of the three-dimensional model of
de Avillez (2000), how the runs used in the current study were set up, and the
global evolution of the ISM as seen in these simulations. In \S~4, the
evolution of the tracer fields and the mixing process in the ISM are
described. Section~5 discusses the results and their implications.
Finally, \S~6 presents a summary of the main results and final
remarks.

\section{Dispersion, Diffusion and Turbulent Mixing in the ISM}

The ISM is a turbulent medium with inhomogeneous makeup driven by SN
explosions.  
The dominance of SN explosions was deduced by Cox \& Smith
(1974) and McKee \& Ostriker (1977) from the detection of hot gas in
UV absorption lines and X-ray emission.  That these explosions
interact to drive a turbulent flow was demonstrated numerically by
Rosen \& Bregman (1995).  The nature of that flow in three
dimensions has been explored with high-resolution hydrodynamical
simulations by de Avillez (2000) and de Avillez \& Berry
(2001), while Korpi et al.\ (1999a, 1999b) included the effect of
magnetic fields, at lower resolution.
The SN explosions individually
and collectively transport hot gas over large distances in more or
less laminar flows.  Turbulence driven by the interactions of these
flows forms eddies that mix the gas on smaller scales.  On the
smallest scales molecular diffusion finally completes the mixing
process.

\subsection{Molecular Diffusion}

Molecular diffusion is the consequence of frequent, stochastically
distributed collisions between gas particles. 
The individual particles scatter in a random
walk around their initial location.  The total squared displacement of
a particle
\begin{equation}
<\Delta x>^{2}=2 D t
\end{equation}
where $D$ is the diffusion coefficient given by 
\begin{equation}
\label{eq9}
D= \lambda u_{\rm rms},
\end{equation}
$u_{\rm rms}$ is the root mean square velocity of the particles, and
$\lambda$ is the mean free path of the particles. The distribution of
the particles over time can be described by a Gaussian distribution with
a standard deviation $\sigma=\sqrt{u_{\rm rms} \lambda t}$ around the
initial location of the particles.

If collisions happen in a homogeneous gas enclosed in a fixed volume,
the relevant quantity to describe the diffusive process is the mean
free path. It does not make sense to talk about a diffusion
coefficient or an expected displacement because, viewed from the
outside, the collisions do not change the properties of the gas, only
the positions of the individual molecules.

The situation is different if there is a composition gradient. Then
there are more particles of the species under study in one part of the
volume than in the other. A random walk carries more particles out
from the higher density region than in from the lower density
region. Net transport results, reducing the gradient and eventually
leading to a homogenized distribution. The streaming ${\mathbf S}$ of
particles can be described as
\begin{equation}
\label{eq10}
{\mathbf S}=-\stackrel{\leftrightarrow}{\mathbf D}\nabla N
\end{equation}
with $\stackrel{\leftrightarrow}{\mathbf D}$ being the diffusion
tensor for anisotropic diffusion and $N$ the number of particles per
unit volume. The gradient is the driving force for the flow, so a
larger gradient leads to a larger flow. The flow also depends on the
mobility of the particles, described by the diffusion tensor. Since
the diffusion tensor depends on the particle speed and on the mean
free path, for a given gradient, the streaming $\mathbf{S}$, as well
as the average displacement $<\Delta x>$, are largest for fast
particles undergoing few collisions (thus having a large mean free
path) and smallest for slow particles undergoing many collisions.

If the diffusion is isotropic, the diffusion tensor reduces to a
diffusion coefficient and the streaming becomes $S=-D\nabla N$. From
the conservation of mass one derives the diffusion equation
\begin{equation}
\label{eq11}
\frac{\partial N}{\partial t}=\nabla\cdot(D\nabla N)
\end{equation}
If the diffusion coefficient is also independent of the spatial
coordinate the equation can be reduced to
\begin{equation}
\label{eq12}
\frac{\partial N}{\partial t}=D \nabla^{2} N.
\end{equation}
This equation shows that the characteristic diffusion time
over a typical length $L$, is 
\begin{equation}
\tau=\frac{L^{2}}{D}=\frac{L^{2}}{\lambda u_{\rm rms}}.
\end{equation}
The time evolution of the particle density being
transported, say along the $x-$direction, is given by
\begin{equation}
\label{sol1}
N(x,t)=\frac{1}{2\sqrt{\pi D t}}\int_{-\infty}^{+\infty}N_{o}(x^{\prime})
e^{-(x-x^{\prime})^{2}/4 D t}dx^{\prime}
\end{equation}
where $N_{o}(x)$ is the initial particle distribution. 

\begin{figure}
\centering
\includegraphics[angle=-90,width=3.3in]{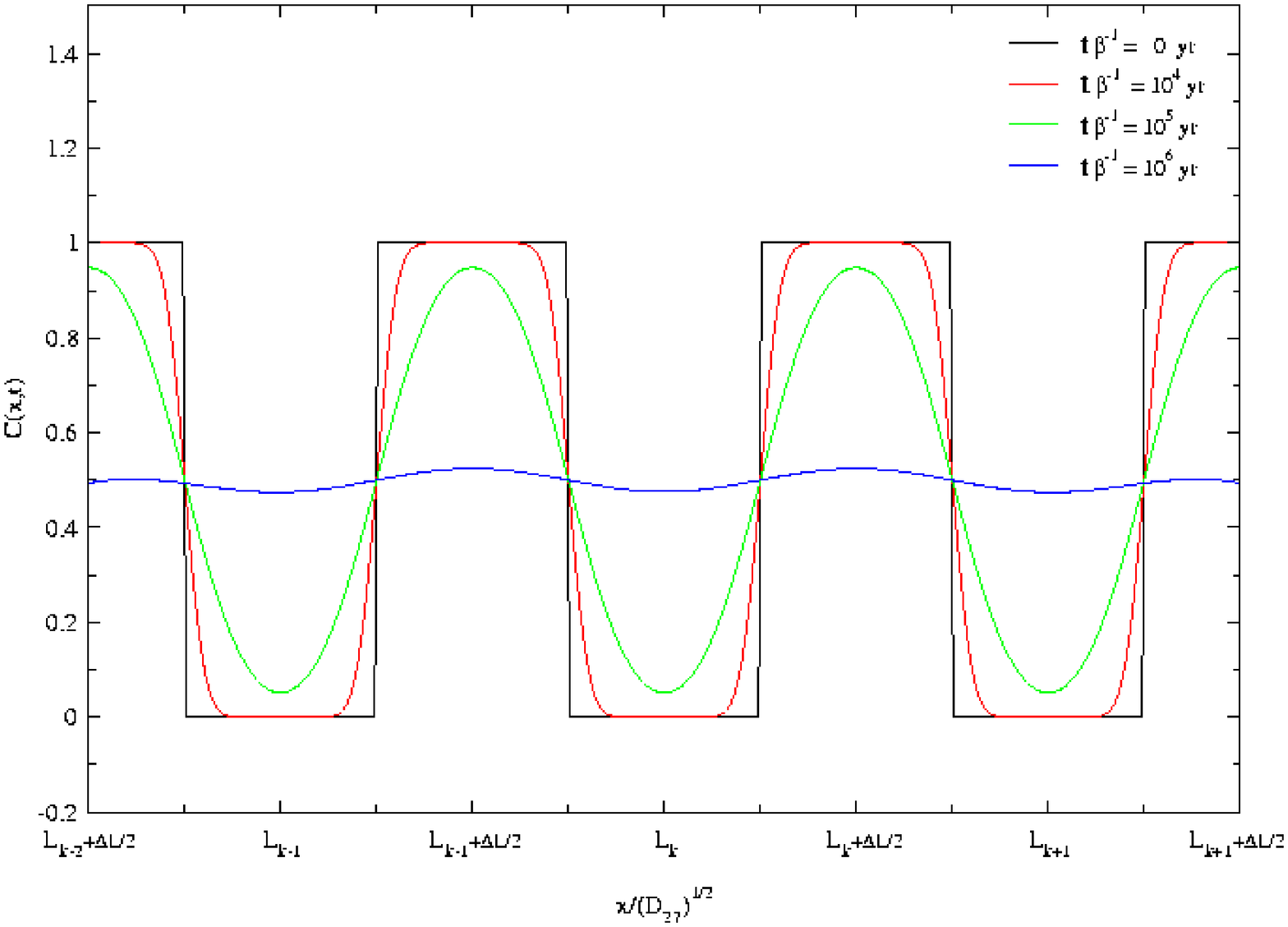}
\caption{Time evolution of the solution (\ref{sol2}) normalized to
$\beta=D_{27}\left(\Delta L/\Delta L_{o}\right)^{2}$. 
\label{errorfunction}}
\end{figure}

\label{sec:solution}
Let us now consider diffusion of an idealized checkerboard pattern
with square size $\Delta L/2$ subject to diffusion with an isotropic,
uniform diffusion coefficient $D$. In one dimension, the pattern can be
described as a square wave in composition $C$ along the $x$-axis with
wavelength $\Delta L$ (see Figure \ref{errorfunction}), so that at $t = 0$,
\begin{equation}
C = C_o(x_{k}^{\prime})=\left\{\begin{array}{lcl} 0 & if & L_{k-1} \le
x_{k}^{\prime}<B_{k-1}^{+} \\

1 & if & B_{k-1}^{+}\le x_{k}^{\prime}< B_{k}^{-} \\

0 & if &  B_{k}^{-}\le x_{k}^{\prime}< L_{k},
\end{array} \right.
\end{equation}
where $B_{k-1}^{+}=L_{k-1}+\Delta L/4$, and $B^{-}_{k}=L_{k}-\Delta
L/4$, The square wave evolves following the solution given in
equation~(\ref{sol1})
\begin{equation}
\label{sol2}
C(x,t)= \\ \frac{1}{2\sqrt{\pi D t}}\displaystyle\sum_{k=-\infty}^{+\infty}
I_k 
\end{equation}
where the solution must satisfy the periodic boundary condition
$C(L_{k-1},t)=C(L_{k},t)$, and the integral $I_k$ has the final form
\begin{equation}
\label{solucao}
\begin{array}{l}
\displaystyle
I_k =  \int_{L_{k-1}}^{L_{k}}C_{o}(x_{k}^{\prime}))\exp\left(
\frac{-(x-x_{k}^{\prime})^{2}}{4 D t}\right)
 dx_{k}^{\prime}=\\ 
 \\
\sqrt{\pi Dt}\left[\erf\left(A\left(x-B_{k-1}^{+}\right)\right)-
 \erf\left(A\left(x-B^{-}_{k}\right)\right)\right],
\end{array}
\end{equation}
where  $\Delta L=L_{k}-L_{k-1}$, $\erf(y)$ is the error function,
\begin{equation}
\erf(y)=\frac{2}{\sqrt{\pi}}\int_{0}^{y}e^{-\eta^{2}}d\eta,
\end{equation}
and if we use a scale of parsecs for $x$ and years for $t$, the value
of the constant $A=8.69/\sqrt{D_{27}t_{\mbox{yr}}}$, with
$D_{27}=D/10^{27}$ cm$^{2}$ s$^{-1}$.  Figure \ref{errorfunction} shows
the time evolution of the solution (\ref{sol2}) for a medium having
inhomogeneities with a scale length of $\Delta L/\Delta L_{o}$, where
$\Delta L_{o}$ is a scale length of reference. The solution is given
in units of $x/\sqrt{D_{27}}$, with $x$ again taken to be in parsecs.

The mixing time scales vary as $(\Delta L/\Delta L_{o})^{2}t_{o}$,
where $t_{o}$ is the time of evolution of a reference square wave with
length $\Delta L_{o}/2=100$ pc. This result is a direct conclusion
from the diffusion timescale. As an example, in a medium with $D=10^{27}$
cm$^{2}$ s$^{-1}$ subject to mixing according to a diffusive mixing
length theory, inhomogeneities with length scale of 100 pc would be
erased in a time of $\sim 10^{6}$ yr, while those with a length scale
of 1 kpc would be erased in a time of $10^{8}$ yr.

\subsection{Turbulent Mixing}

Turbulent diffusion occurs when gas is stirred by more or
less random forces, forming eddies within which
local composition gradients increase, enhancing local molecular
diffusion. 

Studies of isotropic, homogeneous, incompressible turbulence led to
the classical Kolmogorov (1941) cascade picture of turbulent flows.
Eddies with a certain size $\ell$ are assumed to have some typical
velocity $v$ associated with them. The corresponding Reynolds number
$Re = \ell v/\nu$ will be larger for larger eddies, so viscosity $\nu$ is
not very important for them. The energy cascades from these large
eddies to smaller ones. The size of the smallest eddies is set by the
condition $Re = 1$.  Thus, the minimum size of an eddy is
\begin{equation}
\ell_{d}\sim\frac{\nu}{v_{d}},
\end{equation}
and the energy in these eddies is lost to viscous dissipation. Energy
is fed at some rate $\epsilon$ per unit mass per unit time to the
largest eddies of size $\ell_{o}$ and velocity $v_{o}$ with respect to
neighboring eddies, for which the Reynolds number $Re=\ell_{o}v_{o}/\nu \gg
1$. This energy then cascades to smaller and smaller eddies until it
reaches eddies with the smallest size possible, $\ell_{d}$, which
dissipate the energy at the same rate as they are fed in order to
maintain equilibrium. Energy does not accumulate at any scale; the
intermediate eddies merely transmit this energy to the
smaller eddies in a cascade. These intermediate eddies are
characterized by their size $\ell$ and velocity $v$.  Energy passes
through the cascade at a constant rate $\epsilon\sim v^{3}/\ell$,
independent of wavenumber (Kolmogorov 1941). The velocity $v$ of the
eddies scales as
\begin{equation}
v=\epsilon^{1/3}k^{-1/3},
\end{equation}
resulting in an eddy lifetime
\begin{equation}
\tau_{eddy}=\frac{1}{k v}=\epsilon^{-1/3}k^{-2/3},
\end{equation}
where $k\sim 1/\ell$ is the wavenumber. 
Thus, the time taken by an eddy to
disappear varies as $\ell^{2/3}$. An eddy of scale length 
$\ell_{100}$, in units of 100~pc, exists for a time
\begin{equation}
\tau_{100}=(97.8\mbox{ Myr})\frac{\ell_{100}}{v}
\end{equation}
where $v$ is written in units km~s$^{-1}$. We can estimate the time
scale for mixing of hot gas using this classical eddy turnover
time. Taking $v$ as the root mean square velocity of the hot ($u_{\rm
rms}\sim 100$ km~s$^{-1}$) gas and assuming that the eddies have a
length scale $\ell \sim 100$~ pc, the time scale for complete mixing
to occur is roughly 1 Myr.  Therefore, if incompressible turbulent
mixing dominates the mixing process, hot regions will mix very quickly.

\subsection{Dispersal}

Dispersal describes the transport of gas across large regions by
laminar flows, such as those within SN remnants or superbubbles.  On a
large enough scale, encompassing many regions of laminar flow,
dispersion can still be treated as a diffusive process that satisfies
the diffusion equation (\ref{eq12}) with $N$ replaced by the average
density of the ensemble of transported particles $<N({\bf r},t)>$. In
this case it does not make sense to speak about the different phases
of the transported gas, because all of it is dispersed with the flow,
and therefore, the mean free path $\lambda$, of the molecular
diffusion, is replaced by a mixing length path $L$, which is defined
as the distance over which the ensemble of particles is transported
before it halts. Note that it no longer makes sense to define the rms
velocity of each gas component in the transported flow separately,
because the whole ensemble is transported roughly at the same speed.

According to this approximation, the diffusion timescale of a system
with parsec-scale motion can be written as
\begin{equation}
\tau_{\rm dif}=(301\mbox{ yr})\frac{L_{\rm pc}^{2}}{D_{27}},
\end{equation}
where $L_{\rm pc}$ is the length over which the ensemble of particles
is transported, in units of parsecs; $D_{27}=l u_{\rm rms}$ is the
diffusion coefficient, written in units of $10^{27}$ cm$^{2}$ s$^{-1}$;
and $u_{\rm rms}$ is the root mean square velocity of the emsemble of
particles being transported. If one assumes that the characteristic
length scale over which the flow is transported is the same as its
mixing length, then the diffusion time scales as $\tau_{\rm dif}\sim
L/u_{\rm rms}$

\section{Simulations}

\subsection{Supernova-Driven ISM Model}

To study the mixing timescales in the ISM we added tracer fields to
a modified version of the three-dimensional, SN-driven, ISM model
of Avillez (2000). The model is run with an adaptive mesh refinement (AMR)
hydrodynamics code on a $1\times 1\times 20$ kpc region of the
Galactic disk. It includes a fixed gravitational field provided by the
stars in the disk, radiative cooling assuming optically thin gas
in collisional ionization equilibrium, and uniform heating due to
starlight. The radiative cooling function is a tabulated version of
that shown in Figure~2 of Dalgarno
\& McCray (1972) with an ionization fraction of 0.1 at temperatures
below $10^4$~K and a temperature cutoff at 10~K. Background heating
due to starlight varies with $z$ as described in Wolfire et al.\
(1995) and is kept constant in the directions parallel to the plane;
in the Galactic plane at $z=0$ it is chosen to initially balance
radiative cooling at 8000 K. The presence of background heating leads
to the creation of thermally stable phases in the ISM. Due to the
presence of a stable phase at low temperatures, the amount of cold gas
seen in these simulations is larger than that found in the
simulations presented in Avillez (2000).

The interstellar gas is initially distributed in a smooth disk with
the vertical distribution of the cool and warm neutral gas given by
Lockman et al.\ (1986) and summarized in the Dickey \& Lockman (1990)
distribution. In addition, an exponential component representing the
$z-$distribution of the warm ionized gas with a scale-height of 1 kpc
in the Galaxy is included, as described in Reynolds (1987).
The model includes supernovae types Ib+c and II. Supernovae of type Ia
are not included in these simulations because their role is not
important in the mixing process that occurs near the Galactic plane,
although they may be important in the disk-halo interaction due to
their scale-height.

The rates of occurrence of SNe types Ib and Ic in the Galaxy are
$2\times 10^{-3}$ yr$^{-1}$, while those of type~II occur at
$1.2\times 10^{-2}$ yr$^{-1}$ (Cappellaro et al.\ 1997).  We take the
total rate of these SNe in the Galaxy to be $\sigma_{\rm gal} =
1.4\times 10^{-2}$ yr$^{-1}$, corresponding to a rate of one SN every
71 yr. These rates are normalized to the volume of the stellar disk in
which a specific SN type is found: a galactic radius of 12 kpc is
assumed for all stellar disks, but their half thicknesses are assumed
to be twice the scale height of the corresponding SNe distributed in
the field. In particular as SNe Ib+c and II are assumed to have the
same scale height then their stellar disks have the volume
$1.6\times10^{11}$ pc$^{3}$.

SNe of types Ib+c, and II are explicitly set up at the normalized SN
rate, with 40\% of them placed at random locations distributed in an
exponential distribution with a scale height of 90 pc, and 60\% in
locations where previous SNe occurred, representing OB associations,
in a layer with a scale height of 46 pc following the distribution of
the molecular gas in the Galaxy.

The first SNe in associations are chosen to occur in locations where
the current local density is greater than 1 cm$^{-3}$. No density
threshold is used to determine the location where isolated SNe should
occur, because their progenitors drift away from the parental
association and therefore, their site of explosion is not correlated
with the local density.  Similarly, later SNe in associations are no
longer determined by gas density.

The SNe are set up at the beginning of their Sedov phases, with radii
determined by their progenitor masses, which are injected into the
location of the explosion. Type II SNe come from early B stars with
masses $7.7 M_{\odot} \le M\le 15 M_{\odot}$ while type Ib+c SNe have
progenitors with masses $M\ge 15 M_{\odot}$ (Tammann et al.\ 1994). In
this model the maximum mass allowed for an O star is 30
$M_{\odot}$. Avillez (2000) describes in detail the algorithm used to
set up the isolated and clustered SNe during the simulations.

\subsection{Numerical Method and Boundary Conditions}

These simulations use the piecewise-parabolic method of Colella \&
Woodward (1984), a third-order scheme based on a Godunov method
implemented in a dimensionally-split manner (Strange 1968) that relies
on solutions of the Riemann problem in each zone rather than on
artificial viscosity to follow shocks. The Godunov approach is typical
of standard numerical techniques in regions where the solution is
smooth. However, in regions with discontinuities, such as strong
shocks, the Godunov method approximates the solution well by
analytically solving an associated Riemann problem. This is an
idealized problem describing the evolution of a simple pressure jump into
shocks and rarefactions, with a contact discontinuity in
between. Monotonicity constraints ensure that these discontinuities
remain sharp and accurate as they traverse the computational grid. The
higher-order spatial interpolation in the PPM allows steeper
representation of discontinuities.

During the simulation, the mesh is refined periodically in regions
with sharp pressure variations using the AMR scheme.  The local
increase of the number of cells corresponds to an increase in linear
resolution by a factor of two (that is, every refined cell is divided
into eight new cells). At every new grid the procedure outlined above
is carried out, followed by the correction of fluxes between the
refined and coarse grid cells. The adaptive mesh refinement scheme is
based on Berger \& Colella (1989), but the grid generation procedure
follows that described in Bell et al.\ (1994).

The computational domain has an area of 1 kpc$^{2}$ and a vertical
extension of 10 kpc on either side of the midplane.  In the
simulations discussed here, AMR is used in the layer
$\left|z\right|\le 500$ pc. In the highest resolution runs, three
levels of refinement are used, yielding a finest resolution of 1.25
pc. For $\left|z \right|> 500$ pc the resolution is 10 pc. Periodic
boundary conditions are used on the side boundaries, while outflow
boundary conditions are used on the top and bottom boundaries.

\subsection{Tracer Field Implementation and Mixing Runs}

To follow how composition differences mix, we use a tracer
field.  A tracer field acts as a drop of ink in a fluid. In time the
drop will spread and eventually mix completely into the fluid. This
occurs because thermal motion leads to collisions between ink and
fluid molecules, distributing both species uniformly. The simulations
do not include any physical diffusion term, however, so contrary to
what happens in a real fluid or gas, the mixing of the tracer field
occurs by numerical diffusion, which will generally be faster and
larger scale than the physical diffusion in astrophysical problems. As
a consequence the mixing time of a tracer field in our model provides
a rather strong {\em lower} limit for the timescale of mixing
resulting from physical diffusion.

The tracer field is an intensive scalar whose advection is computed
with the same algorithm as the density.  We initially set its value to
be either one or zero to represent regions of different
composition. In the absence of diffusion, either numerical or
physical, these values would remain fixed as the gas is advected. In
the presence of diffusion, the tracer field values will change.  In
our simulations, this is a result of the numerical diffusion present
in the Godunov advection scheme.  

This scheme assumes that the flow solution is represented by a series
of piecewise constant states, and therefore, that the numerical
representation approximates the true solution near
discontinuities. The numerical solution is evolved by considering the
nonlinear interaction between these piecewise constant states.  Viewed
in isolation, each pair of neighbouring states constitutes a Riemann
problem. The algebraic solution in the overall grid results from a
collection of Riemann problems for all the interfaces between two
sucessive cells at every time step. The discontinuity between two
states evolves into some combination of shock and rarefaction waves,
separated by a contact discontinuity (Figure \ref{riemann}). The
tracer field is advected in the same way as the density and therefore,
it diffuses between any rarefaction wave and the contact discontinuity
formed as a result of the Riemann problem solution. This process
happens between every pair of neighboring cells, leading sooner or
later to a change of the tracer field value.  In the case at hand we
use tracer fields to follow the evolution and mixing of composition
differences in the ISM.  As mixing occurs by numerical diffusion,
regions with values intermediate between zero and unity grow, until
ultimately the entire volume is filled with gas having value roughly
0.5.

The tracer field is set up initially with values of $C=0$ and~1 on
alternating squares of a checkerboard in the plane of the Galaxy with
square sides of $l=25,~50$ or 500~pc (Figure 2). The 500~pc scale
corresponds to a tracer field that fills a quarter of a 1 kpc length
square board.  The distribution is uniform in the vertical direction;
that is, the squares are extended into rectangular solids of uniform
composition vertically.  This setup allows a direct comparison to the
analytic solution given in \S~\ref{sec:solution} for the mixing of
inhomogeneities of different length scales in the ISM due to classical
turbulent diffusion.

In the current work we report on simulations using seven SN rates
$\sigma / \sigma_{\rm gal} =1$, 5, 10, 15, 20, 30 and 50, having
finest grid resolutions of 1.25, 2.5 or 5 pc. In general, the
simulations were stopped between 50 and 70 Myr after complete mixing
occurred. The simulations ran for 400 Myr for $\sigma/\sigma_{\rm gal}
=1$, 300 Myr for $\sigma/\sigma_{\rm gal} =5$, and 200 Myr for the
remaining SN rates. A summary of these runs is presented in
Table 1.

\section{Results}

\subsection{Overview of Results}

In the first 50 Myr the system evolves into a statistical steady state
on the global scale.  Once disrupted by SN explosions, the disk never
returns to its initial state, provided SNe continue to
explode. Instead, regardless of the initial vertical distribution of
the disk gas, a thin disk of cold gas forms in the Galactic plane,
and, above and below, a thick inhomogeneous gas disk forms.  Gas flows
between the thin and thick gas disks, with upward and downward flowing
gas coming into dynamical equilibrium.  The upper parts of the thick
disk form the disk-halo interface, where a large scale fountain into
the halo is driven by hot ionized gas escaping in a turbulent
convective flow from the disk-halo interface and from superbubbles
blowing out of outer layers of the thin disk.

Table 1 gives the time for complete mixing, measured in the supernova
forming stellar disk, which we take to be the time needed for the
average value of the tracer field to saturate at a value $\sim 0.5$
(see \S4.2). Results are shown for checkerboards with initial square
sizes of $l=25$, 50, and 500 pc, and SN rates
$\sigma$/$\sigma_{gal}=1$, 5, 10, 15, 20, 30, and 50, at resolutions
$\Delta x$ of the finest grid levels used during adaptive mesh
refinement.  A quick look at the table shows that the time required
for complete mixing of inhomogeneities with $l=25$ and $50$ pc differs
by only 5--20 Myr, while that for inhomogeneities with $l=500$ pc is
larger by, at most, another 20 Myr. These differences decrease with
increasing SN rate. Runs with different finest resolution $\Delta x$
for the same SN rate show similar mixing time scales, suggesting that
the results depend quite weakly on the actual strength of the
diffusion.

Figures \ref{topview1} and \ref{topview2} present snapshots of the
evolution of tracer fields in checkerboards with initial length scale
of $l=50$~pc and SN rates $\sigma/\sigma_{\rm gal}=1$, 5 and 15.  The
snapshots in Figure \ref{topview1} were taken at 50 and 126.6 Myr,
while those in Figure \ref{topview2} were taken at 50 Myr of
evolution. The resolution of these snapshots is 1.25~pc in
Figure~\ref{topview1} and 2.5 pc in Figure~\ref{topview2}. We see that
for the Galactic SN rate, the mixing process at 50 Myr is just
starting to occur, while for larger SN rates, it is well advanced,
with a larger fraction of the inhomogeneities reduced to weak
structures that take a long time to be erased.  For example, in the
model with $\sigma/\sigma_{\rm gal}=15$ at a time of 50 Myr, the
tracer field has a value varying between 0.4 and 0.6. For the other
rates shown in the figures at 50 Myr there are still unmixed regions
where tracer field has values of zero or one.

These Figures show large-scale laminar flow in the hot gas bubbles
that transports tracer field over large distances.  The expansion and
interactions of the bubble drive shocks into the warm and cold gas.
Vorticity generated at shock intersections drives eddies mixing the
gas up to 100~pc scales, as well as Rayleigh-Taylor fingers that
degenerate into smaller eddies at the tips of their caps. The eddies
cascade from intermediate to small scales, further mixing the tracer
field down to the diffusion scale.  The mixing process thus proceeds
by different mechanisms at different scales, because of the
inhomogeneous, compressible nature of the flow.

The mixing process is independent of the spatial distribution of
isolated SNe that occur randomly in the field. The locations of type
II and Ib+c SNe occurring away from their parental OB associations are
determined from a file of random numbers chosen at the beginning of a
simulation. We ran simulations with different sets of random numbers
and finest grid resolutions of $\Delta x=1.25$ and 2.5 pc, and found
mixing time scales virtually the same as the ones reported here, so we
do not further discuss those models.

\subsection{Average, Variance, and Maximum}
The simulations start with exactly the same number of cells having
value 1 and 0. Therefore, the average value of the tracer field,
$<C>=0.5$ at $t=0$. By definition complete mixing
occurrs when the tracer field at any point in the grid has
$C_{max}=C_{min} = 0.5$. However, due to round-off and phase errors of
the numerical scheme, the actual value for completely mixed material
may differ from 0.5 by a few percent.  Once the tracer field has this
constant value on the entire grid, no further variation occurs.

Figure \ref{averagec} shows the time variation of $<C>$ for three
SN~rates $\sigma/\sigma_{\rm gal}=1,~5$ and~15, and for two finest
grid resolutions of 2.5 and 5 pc. During the first 50 Myr, $<C>$
shows small variations around 0.5, approaching a constant value as
mixing increases and the small scale structure dominates the mixing
process. The initial variations of $<C>$ result from the increased
error occurring as steep color gradients interact with steep velocity
gradients in SN-driven shock waves.  As the color is mixed, the color
gradients decrease, and $<C>$ approaches a constant value close to
0.5.  The biggest errors, still only 2.4\%, occur for the highest SN
rates, where strong shocks are more widely distributed.

The average value of $C$ gives an indication of the time at which
complete mixing occurred, but it does not provide any information of
the mixing process. Further information may be obtained from the
variance of the tracer field, $\sigma^{2}_{C}$.  This evolves from
$\sigma^{2}_{C} = 0.25$ at $t=0$ to $\sigma^{2}_{C} = 0$ when complete
mixing occurs. The variance has a steady decrease with time until it
reaches a value of $10^{-3}$. For $\sigma^{2}_{C} < 10^{-4}$ the
decrease is no longer steady, but rather sporadic, suggesting that we
have reached a regime where numerical noise is dominating.
Figure \ref{variance1} shows the variance of $C$ (for
$\sigma^{2}_{C} \geq 10^{-4}$) over time for inhomogeneities with
different length scales, along with exponential fits to the steady
decrease of $\sigma^{2}_{C}$ until it reaches a value of
$10^{-3}$. The fits shown are for the shorter length scales of the
inhomogeneities. We find that the variance of the tracer field for
inhomogeneities with length scales $l=25$ and 50~pc decays to
$10^{-3}$ remarkably close to exponentially, with $\sigma_{C}^{2}\sim
e^{-t/\tau}$. For $\sigma^{2}_{C} < 10^{-3}$ such decrease is seen for
$\sigma/\sigma_{Galaxy}=1$ or~5 if $l=25$ pc and
$\sigma/\sigma_{Galaxy}\leq 10$ if $l=50$ pc.

Figure \ref{variance2} shows the variation of the time constant $\tau$
with SN rate for inhomogeneities with initial $l=25$ and 50~pc, as
well as the best power-law fits to the data points. The fits to an
inverse square root law are remarkably good, with correlation
coefficients of 0.997 and 0.998 for $l=25$ and 50 pc, respectively.
For length scale of 50~pc,
\begin{equation}
\tau=(37.9 \mbox{ Myr}) (\sigma/\sigma_{\rm gal})^{-0.5}.
\end{equation}
In future work we will extend our parameter study sufficiently to
define the dependence on inhomogeneity length scale $l$ fully.  

This exponential decay law indicates that large variations in the
tracer field decay quickly while smaller variations last much
longer. Such a decay law for the variance of $C$ breaks down for
larger inhomogeneities, and in particular it does not apply for
inhomogeneities with $l=500$ pc.  The extended parameter study should
show at which length scale the deviation from this behavior begins to
occur.  We will also examine whether this result extends
to smaller scales using finer grids. 

The time variation of the maximum value of the tracer field, $C_{max}$
is shown in Figure \ref{cmaxtime}.  The exponential decay behavior is
again seen, with a fast initial decline followed by a slow approach to
complete mixing.  Structures with small variations in composition will
thus be present for extended periods of time. The figure also shows
that the time taken for complete mixing to occur decreases with
increasing SN rate until it saturates for $\sigma/\sigma_{\rm gal}>
20$. This result is also seen in the variation of the mixing time with
SN rate shown in Figure \ref{mixtimesnrate}, which shows that for SN
rates $\sigma/\sigma_{\rm gal}> 20$ the mixing time scales differ by
only a few Myr.

The mixing process has a rather weak dependence on the resolution used
in the calculations, and thus of the actual strength and scale of
diffusion. Running the scheme with a better resolution decreases the
numerical errors, and therefore also the magnitude of
numerical diffusion. Therefore, the mixing times become longer (going
from right to left column in Table 2), but this increase is far
slower than the increase in resolution, and declines with increasing
resolution, as shown in Table 2 and Figures~\ref{mixtimeresolution}
and~\ref{mixtimesnrate}.  This demonstrates that numerical convergence
is occurring, and further that unresolved details of the mixing
process at small scales do not affect the behavior at resolved
scales. This appears to be because the mixing process depends
primarily on motions near the SN driving scale of $l\ge 100$ pc, with
the presence of small scale diffusion more important than its exact
properties.

The similarity in the mixing time scales for $\sigma/\sigma_{\rm
gal}> 20$ occurs because for SN~rates higher than this the entire disk
quickly heats.  Figure~\ref{PDF-T} shows the temperature probability
distribution functions (PDFs) of the disk gas for $\sigma/\sigma_{\rm
gal}=20$. After the first 50 Myr of evolution a large fraction of the
disk gas already has temperature $T> 10^{5}$~K and by 100 Myr of
evolution all the gas has reached at least $10^{5}$~K. As a
consequence of the high SN rate the disk becomes warmer and cooling
becomes inefficient. The high rms velocities in this hot gas lead to
quick mixing, as in hot regions in less energetic models.  The
complete lack of colder regions, however, means that all the gas mixes
on the same short time scale. Once there is no further cold gas,
increasing the SN~rate further only slowly increases the rms velocity,
leading to the saturation behavior seen.

We emphasize that classical mixing length theory badly fails to
predict the mixing time scales that we compute. That theory predicts
that the mixing time scales should depend on the length scales.  That
is, let a checkerboard be composed of squares with length scale $L$,
and the squares be further divided into $n$ smaller squares with length
scale $l$. Then mixing length theory would predict the diffusion time
scales of these squares to be related by $\tau_{L}/\tau_{l}=n^{2}$ for
constant diffusion coefficient. Thus, the predicted ratios of
timescales (for example for $L$ and $l$ in parsecs) 
are $\tau_{500}/\tau_{50}=100$
and $\tau_{50}/\tau_{25}=4$.  However, Table~1 shows that the actual
ratioes are never greater than 1.2, a value far smaller than predicted.

\subsection{Profiles}

Another approach to measuring the mixing timescale is to directly
compute the diffusion coefficient required to fit the average profile
across an initially sharp color gradient, using the analytic theory of
\S~\ref{sec:solution}. Figure \ref{fitD1} presents profiles 
of the tracer field averaged along the $x$-axis for $y\geq 500$ pc for
the checkerboard with squares of initial length of 500 pc (shown in
the top panel of Figure 2), overlaid by the classical diffusion
solution from equation~(\ref{solucao}) with $n=1$. Initially the
solutions fit the average profile of the tracer field well, but at
later times the fits become somewhat worse. The correlation factors of
these fits reduces as time evolves from 0.998 at 25 Myr to 0.94 at 225
Myr.

From the fits, we can deduce the diffusion coefficient $D_{27}$ written
in units of $10^{27}$ cm$^{2}$ s$^{-1}$. The time variation of
D$_{27}$ for two different SN~rates is shown in Figure \ref{fitD}. The
straight lines are the best fit to the data points.  We find that the
diffusion coefficient varies exponentially with time, with the general
expression of
\begin{equation}
\log D_{27}=A+t/\tau_D,
\end{equation}
where, if $t$ is given in Myr, $A=-2.91$ and $\tau_D=107$ Myr for
$\sigma/\sigma_{\rm gal}=1$, and $A=-2.45$ and $\tau_D=67.1$ Myr for
$\sigma/\sigma_{\rm gal}=5$. If the mixing were regulated by classical
diffusion, then we would expect that the diffusion coefficient would
be nearly constant or vary around a mean value with time, rather than
exponentially increasing.

\section{Discussion}

In this paper we try to understand how long it takes to erase
inhomogeneities in the ISM and if this process depends on the
inhomogeneity length scale. That is, does it take longer to mix big
regions than small ones? We modified the three-dimensional model of
Avillez (2000) to incorporate tracer fields and ran simulations with
different SN rates, distributions of SNe in the field, and numerical
resolutions. (Clustered SNe from OB associations are explicitly set
up, with the first SN in an association occurring in a location
where the local density is greater than 1 cm$^{-3}$.)

The three-dimensional adaptive mesh refinement code used in our
simulations, unlike 1-zone chemical evolution models (Matteucci et
al.\ 1999; Tosi et al.\ 1998) and chemodynamical models (Samland et
al.\ 1997; Samland 1998; Argast et al. 2000, to name a few), enables us to
resolve local inhomogeneities in the ISM with a spatial resolution as
good as 1.25~pc. This allows us to test the usual assumption of
Galactic chemical evolution models that assume instantaneous mixing of
the ISM at all times.  Mixing in the real world takes finite time, and
ultimately occurs on the molecular scale, through physical
diffusion. We set a lower limit on the mixing time scale by assuming
that it is set at the far larger grid scale by faster numerical
diffusion.  We find that the actual value of the mixing scale does not
strongly influence the mixing timescale.

Matter can enter and leave the section of the Galaxy under study
through the top and bottom outflow boundary conditions, as well as
moving to neighboring periodic boxes (eg passing through a periodic
side boundary condition), although we do not yet include Galactic
rotation in our model.  Chemical evolution models do not normally
allow radial transport of matter between different rings, 
but it occurs in our models.

Because the domain extends to 10 kpc on both sides of the midplane,
gas can cycle between the disk and the halo in a Galactic fountain as
a result of the dynamics of the flow in the gravitational field of the
Galaxy (see Kahn 1981; Avillez 1999).  The mixing time scales
of the disk gas are regulated mainly by the SNe, because the
disk gas that enters the disk-halo cycle takes some 200 Myr to return
to the disk (Kahn 1981). Furthermore before reaching the
thin gas disk, the descending gas hits the thick gas disk and
decelerates due to drag (Kahn 1981; Benjamin \& Danly
1997). Therefore, most of the descending gas will not contribute
strongly to the mixing process of the ISM with $\left|z\right|\leq
250$ pc.

This work only studies the mixing process as a way of erasing
inhomogeneities in the ISM. It does not take into account continuing
pollution of the ISM by SNe, planetary nebulae, and stellar winds,
which we will study in future work.  The actual enrichment of the disk
gas depends on the sources and the efficiency of mixing.  If the
ejected metals are quickly distributed over a large volume, spatially
homogeneous enrichment takes place. If the mixing volume is small, the
ISM in the vicinity of a SN will be highly enriched, while large parts
of the disk remain metal-poor until the gas mixes out on time scales of 
hundreds of Myr or more. In this case the ISM may remain chemically
inhomogeneous, with newly formed stars having different compositions
depending on where they form.  Over gigayear timescales, mixing does appear 
to be quite effective over scales of many kiloparsecs.

This work shows that inhomogeneities that are present in the ISM,
assuming that no further inhomogeneities are introduced into the
system, can take a long time to be erased. If inhomogeneities are
introduced at a time larger than the time scale for mixing, the system
will show temporarily an homogeneous state, which ends when
inhomogeneities are introduced. On the other hand, if inhomogeneities
are introduced on a time scale smaller than the mixing time, then the
system will show an inhomogeneous distribution during its
evolution. 
Therefore, from the simulations we conclude that for the
ISM to show a homogeneous distribution, no inhomogeneities should be
introduced over periods of roughly a hundred megayears.  This is
longer than typical star formation time scales, so new SNe and
pre-main sequence stars can continue to maintain an inhomogeneous
composition so long as star formation continues.

The rate at which inhomogeneities are erased increases with the
SN rate to values about ten times greater than the Galactic
rate.  Above that rate, no cold medium remains, mixing occurs quite quickly, 
but above this threshold, further increasing the SN rate does not markedly 
further increase the mixing efficiency.

The time for complete mixing is already longer than any time scale for
chemical equilibrium to occur and therefore the ISM will show an
uneven distribution. Even if the major fraction of the ISM gas is
completely mixed, 
lower level inhomogeneities will still be seen as abundances are
measured more carefully.
Furthermore, as the mixing time scales are
longer than the time interval for SNe type II to occur and 
enrich the ISM with metals, low-level inhomogeneities are continuously 
maintained,
leading to a poorly mixed ISM, except in the hot regions where the
mixing process is fast (of order 50~Myr).

For a galactic SN rate, the ISM is likely fairly inhomogeneous, since
the SNe are spatially well separated and erasing these inhomogeneities
takes a long time (of the order of 350 Myr). A single SN event can
thus substantially influence its surroundings.  If we consider, for
example, a 20~pc sphere with number density $n=1$ around a SN with
1~M$_{\odot}$ of metals in its ejecta, the ejecta represent a fraction
$\sim 0.1 (Z/Z_{\odot})$ of the metals in the region.  Thus,
individual SNe can drastically change the local metallicity in low
metallicity gas.  Stars formed before mixing has finished can have a
wide range of metallicities; further work will be needed to make
quantitative predictions of the expected dispersion for comparison with
observations, however.

The smallest mixing time scale that we find in our study is $\sim 120$
Myr and occurs for $\sigma/\sigma_{\rm gal}\ge 20$.  After some $80\%$
of the gas has been mixed, the simulations show that the ISM remains
inhomogeneous for at least twice as long on scales smaller than a
kpc. Even when the rate of SNe is increased to ten times the Galactic
rate, this uneven abundance distribution structure takes tens of Myr
to disappear.

At the finest grid scale we suppress the turbulent diffusivity, as
numerical viscosity becomes the dominant diffusion mechanism.  At
these scales there would physically be a continuing turbulent mixing
that we do not resolve. The resolution study shown in
Figures~\ref{mixtimesnrate} and~\ref{mixtimeresolution} demonstrates,
however, that the mixing time is practically independent of the actual
value of the diffusion scale.  Changing the linear diffusion scale by
factors of as much as four produced less than 20\% changes in the
diffusion time.  This means that we can effectively assume that the
mixing is dominated by an inertial scale decoupled from the diffusion
scale, so that the details of how the diffusion occurs do not affect
the larger-scale behavior.

\section{Summary and Final Remarks}

The main results of this study are:
\begin{itemize}
\item The mixing of a SN-driven ISM is characterized by three regimes:
laminar flows dominate on scales close to a kpc, turbulent mixing
dominates below about a hundred parsecs, and diffusive processes
finally dominate at the smallest scale.
\item Mixing at the larger scales appears nearly independent of the
exact value of the diffusive scale, so long as some diffusive process
ultimately operates at small scales.
\item The time scale to erase inhomogeneities in the ISM on length
scales from parsecs to kiloparsecs is nearly independent of the length
scale; classical mixing length theories badly fail to describe this
behavior. 
\item The time scale for complete mixing, providing no further
inhomogeneities are introduced, is some 350 Myr for
the Galactic SN rate, decreasing with increasing SN 
rate up to a threshold value.  Above this threshold value, further increases
in the SN rate have little further effect.
\item The speed of mixing decreases almost exponentially, with a time
constant dependent on the inverse square root of the SN rate: strong
inhomogeities decay quickly, while weaker ones last much longer.
\item The diffusion coefficient derived from fits of a solution of the
diffusion equation to the simulation results does not remain constant as 
assumed by classical mixing length theory, but rather increases exponentially
with time, with a time constant dependent on the SN
rate.
\end{itemize}

This work is far from complete.  At least four issues clearly remain
to be addressed.  First, we need to determine the actual length scale
at which the exponential decrease in tracer field variance ceases,
somewhere between 50 and~500~pc.  Second, if further inhomogeneities
are introduced, such as the chemical elements fed into the ISM by SNe,
planetary nebulae, and stellar winds, large parts of the ISM, or
indeed the whole ISM, may never reach homogeneity.  Even though
stronger inhomogeneities decay quickly, the exponential decline in
mixing rate for weaker inhomogeneities means that a wealth of weaker
inhomogeneities will remain.  Determining quantitative values of the
dispersion of metallicities will help solve outstanding questions in
studies of stellar populations and the ISM. Third, the simulations do
not include the Galactic magnetic field. One can speculate that the
presence of magnetic field would contribute to an increase of the
mixing time scales in the Galaxy. It could reduce or suppress
turbulent flows, and also it would reduce the expansion of laminar
flows. Furthermore, the random component of the field could enhance
the diffusion in the SN shells, thereby fragmenting the shells more
rapidly and affecting the mixing process. We plan to run a new set of
simulations to determine dependence of the mixing process on field
intensity and distribution.  Finally, a theoretical approach,
supported by direct numerical simulations, must be developed in order
to explain the mixing process in the ISM, because as this work shows,
the classical approach to the problem fails.

\acknowledgements

The authors thank D. York, E. Jenkins, and R. Ferlet for emphasizing
the importance of this problem and them and M. Saha for useful
discussions. This work was supported by an NSF CAREER grant (AST
99-85392), and has made use of the NASA Astrophysical Data System.

\clearpage

\begin{figure}
\centering
\includegraphics[angle=-90,width=2.5in]{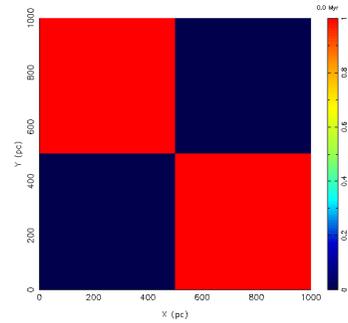}
\includegraphics[angle=-90,width=2.5in]{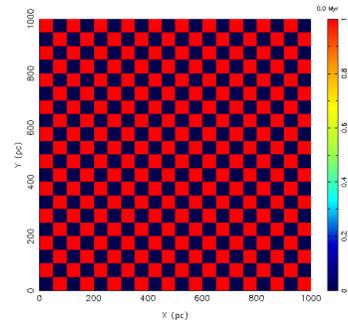}
\includegraphics[angle=-90,width=2.5in]{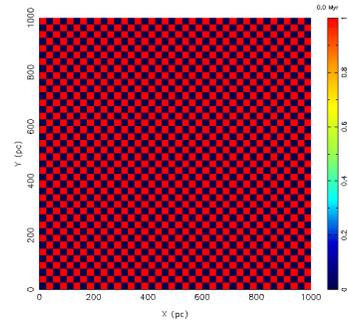}
\caption{Tracer field set up as checkerboards with squares of $l=500$ (upper panel), 50 (middle panel) and 25 pc (lower panel). The red squares have value 1, while the blue regions have the value 0.
\label{tracersetup}}
\end{figure}

\begin{figure}
\centering
\includegraphics[angle=0,width=3in]{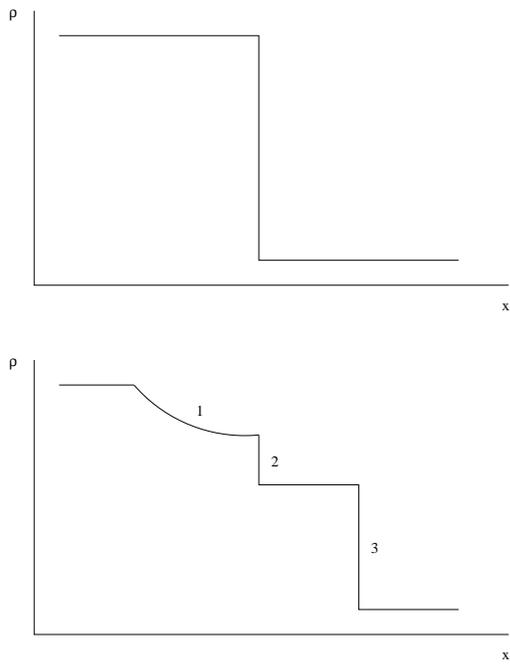}
\caption{Solution to the Riemann problem between two neighboring cells. The figure shows the density profile along the $x-$direction of the Riemann problem before (left) and after (right) the problem is solved. The right plot shows the location of the rarefaction wave (1), contact discontinuity (2), and forwards shock wave (3).
\label{riemann}}
\end{figure}

\clearpage 

\begin{figure*}
\centering
\includegraphics[angle=-90,width=3.4in]{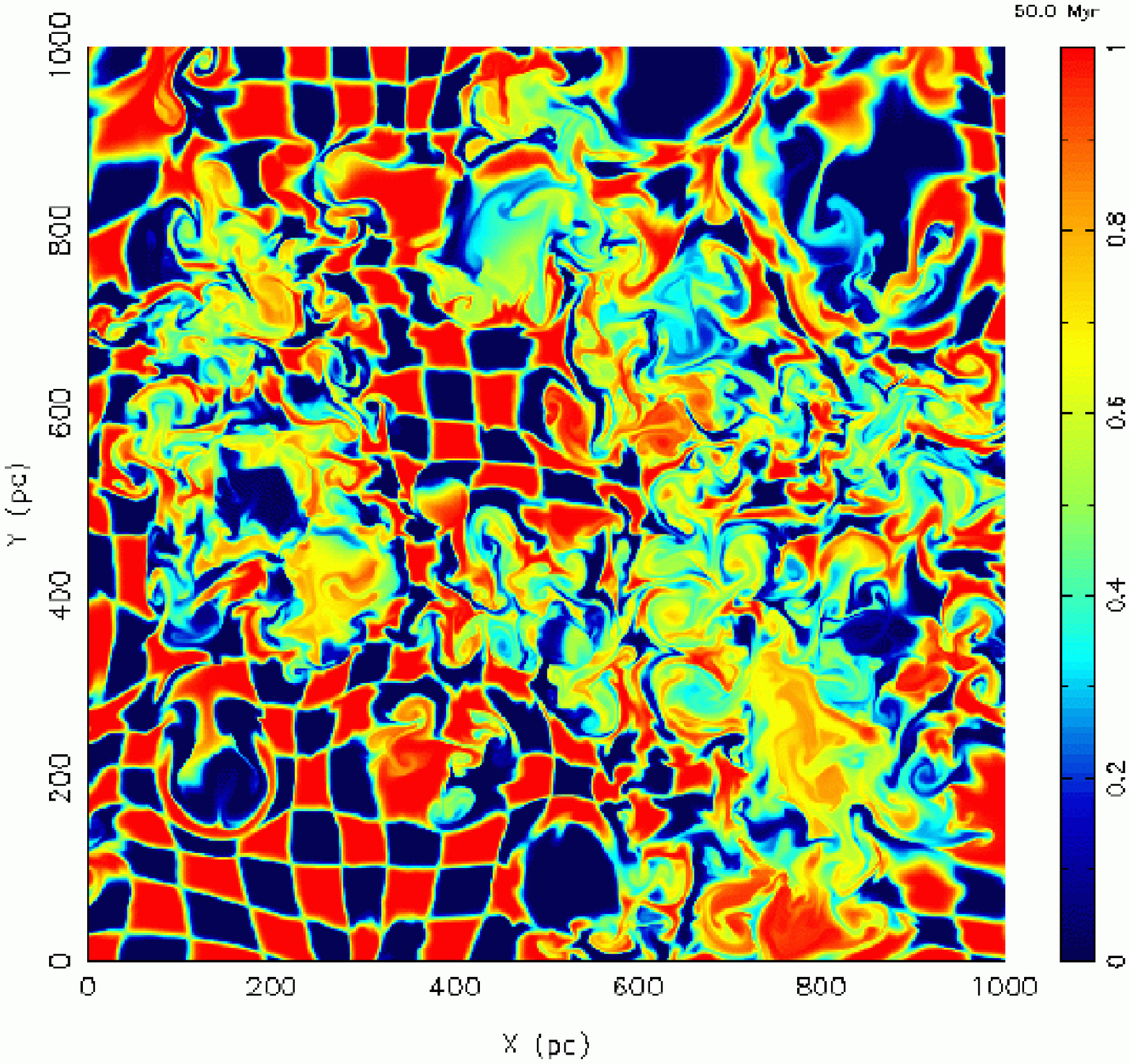}\includegraphics[angle=-90,width=3.4in]{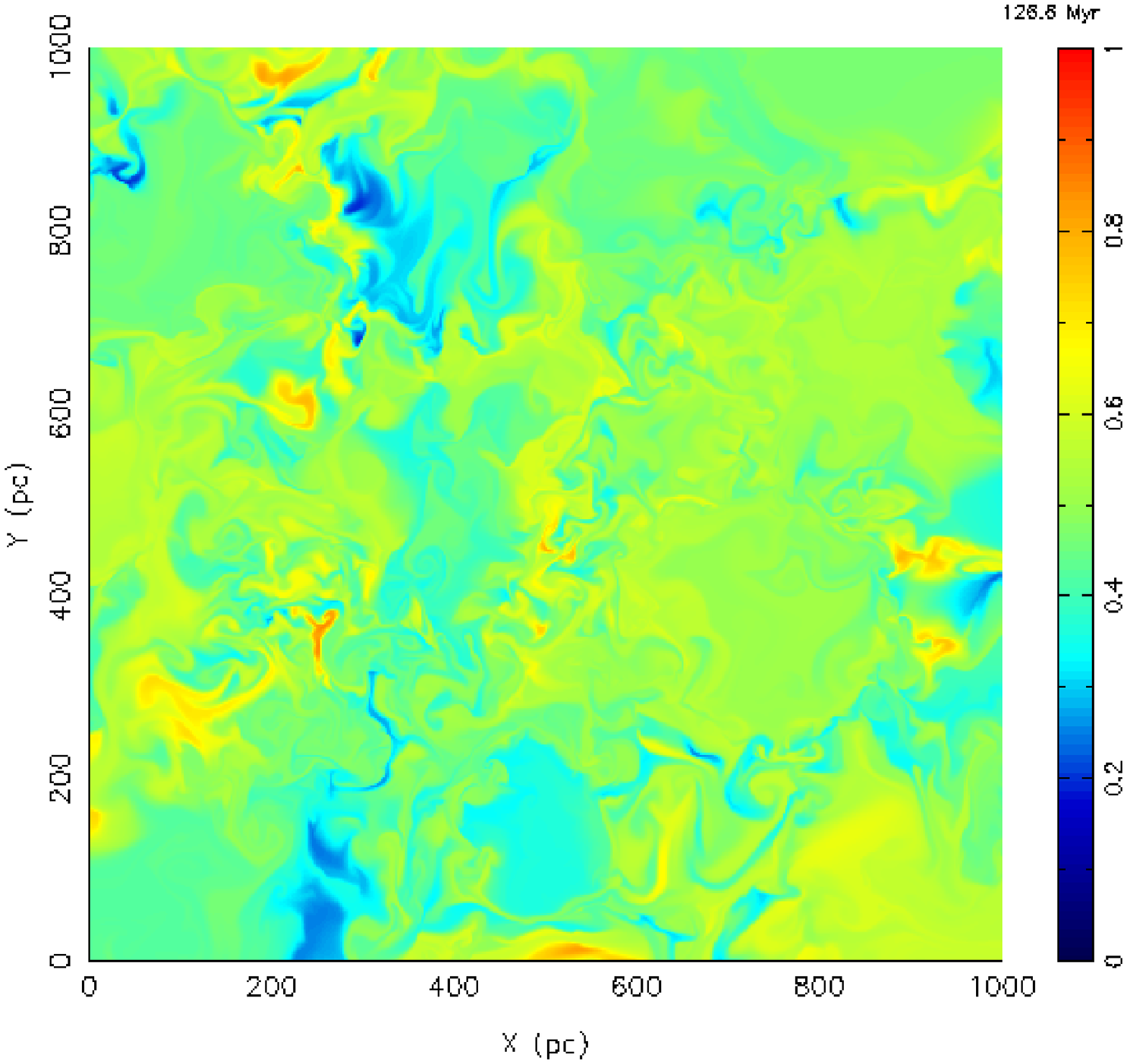}
\caption{Distribution of inhomogeneities with length scale $l=50$ pc at times 50 and 126.6 Myr of evolution for the Galactic SN rate. The resolution of these images is 1.25 pc.}
\label{topview1}
\end{figure*}

\begin{figure*}
\centering
\includegraphics[angle=-90,width=3.4in]{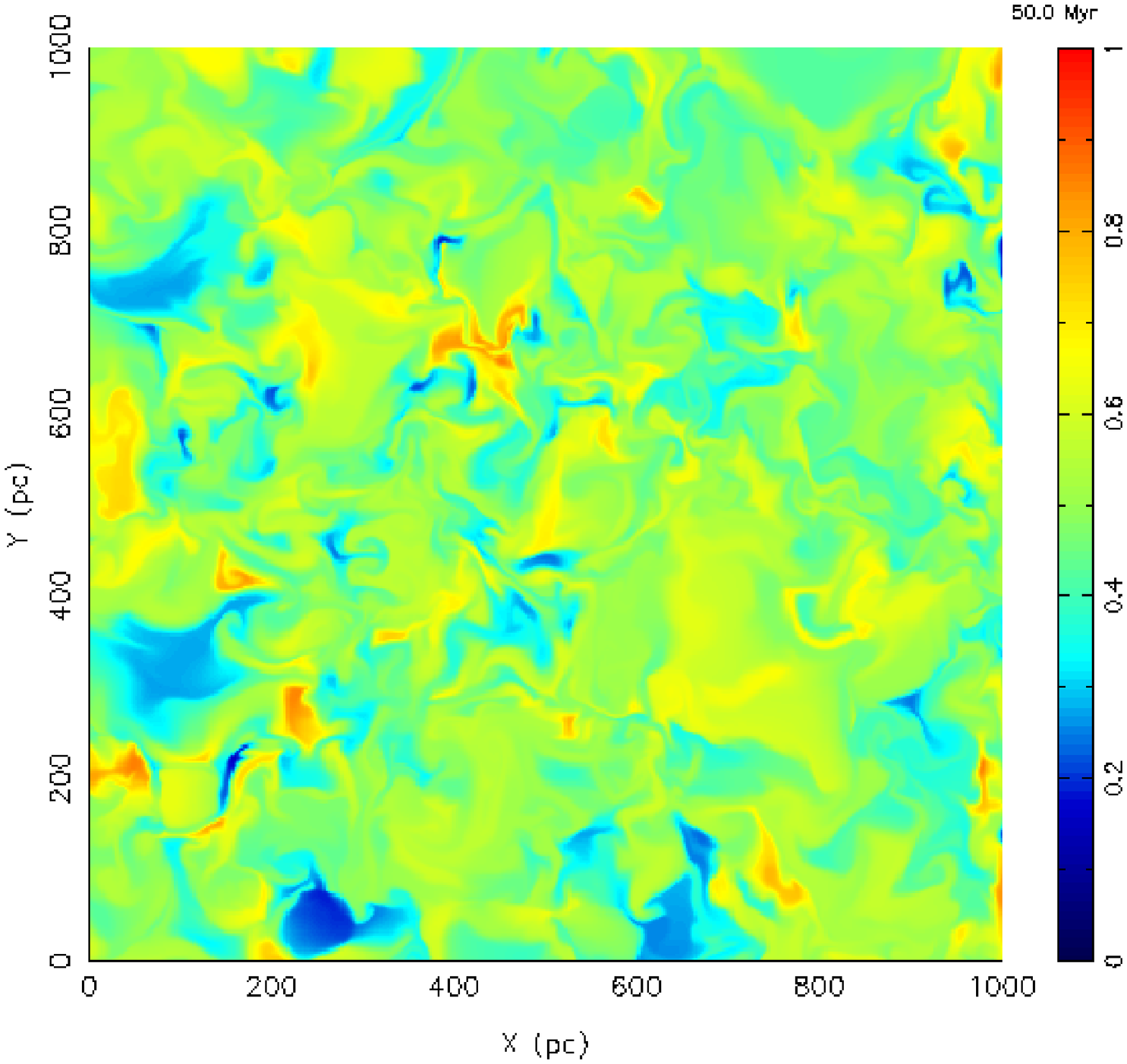}\includegraphics[angle=-90,width=3.4in]{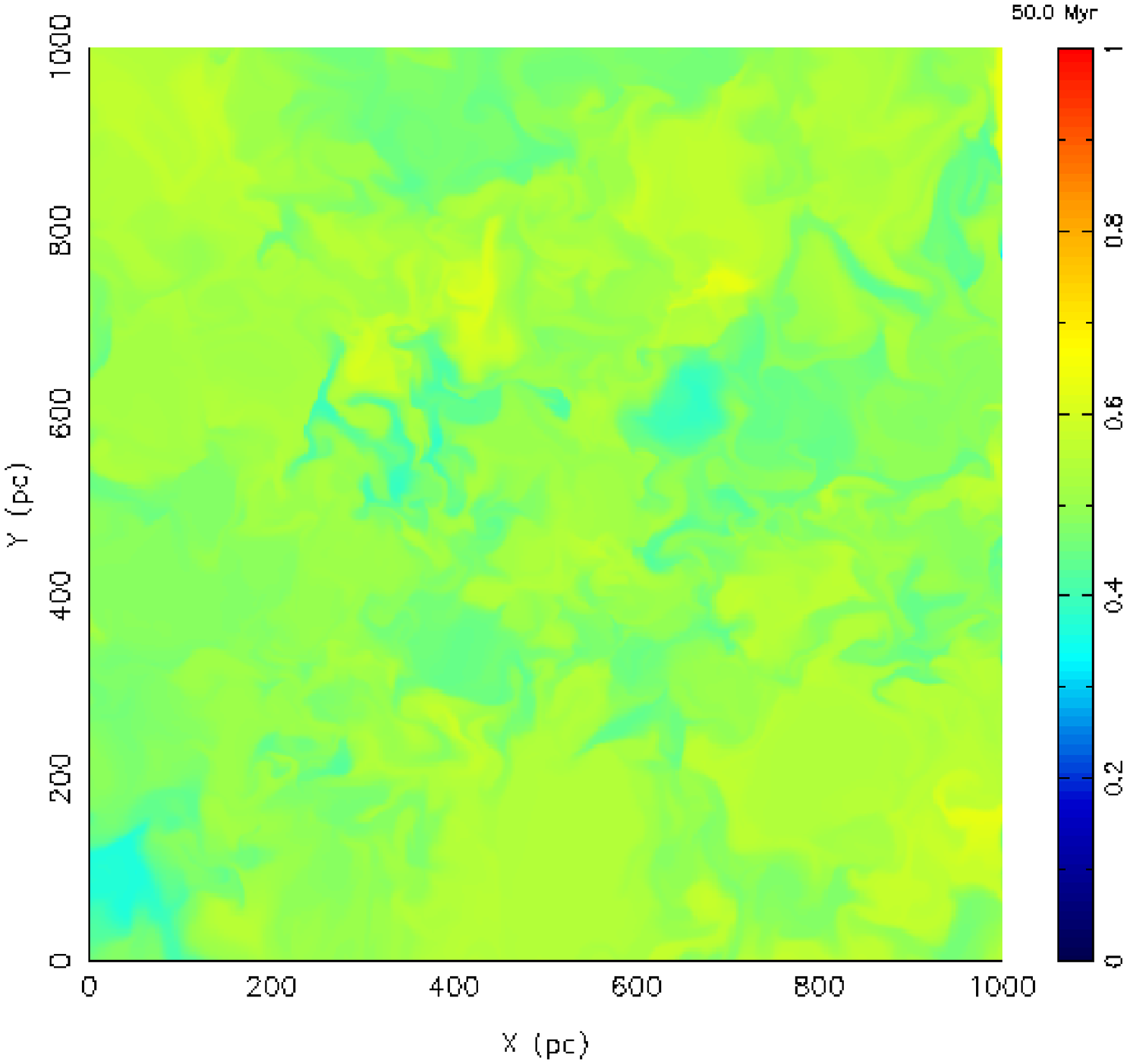}
\caption{Comparison between the mixing process at 50 Myr of evolution in an ISM where the SN rate is 5 (left panel) and 15 (right panel) times the Galactic rate. Note that for $\sigma/\sigma_{\rm gal}=15$ after 50 Myr of evolution most of the inhomogeneities have been mixed, remaining only the small scale structure which takes a long time to mix. The resolution of these images is 2.5 pc.}
\label{topview2}
\end{figure*}

\clearpage

\begin{figure}
\centering
\includegraphics[angle=-90,width=3.3in]{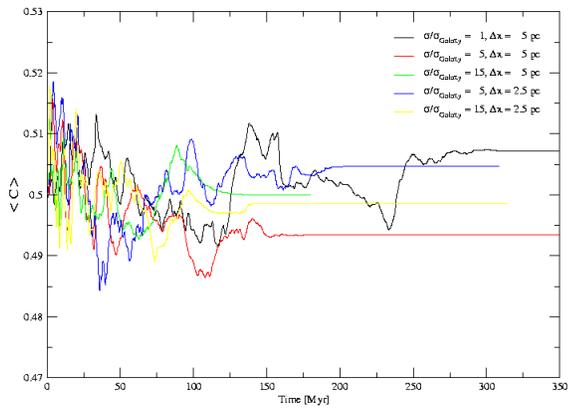}
\caption{Time variation of $<C>$ in the checkerboard for three rates $\sigma/\sigma_{\rm gal}=1,~5,~15$ for finer level resolutions of 2.5 and 5 pc. The larger deviations of $<C>$ from 0.5, when complete mixing occurred, are $1.32-1.4\%$ ocurring for $\sigma/\sigma_{\rm gal}=1$ and 5, respectively, at a resolution of 2.5 pc.}
\label{averagec}
\end{figure}
\clearpage
\begin{figure}
\centering
\includegraphics[angle=-90,width=3in]{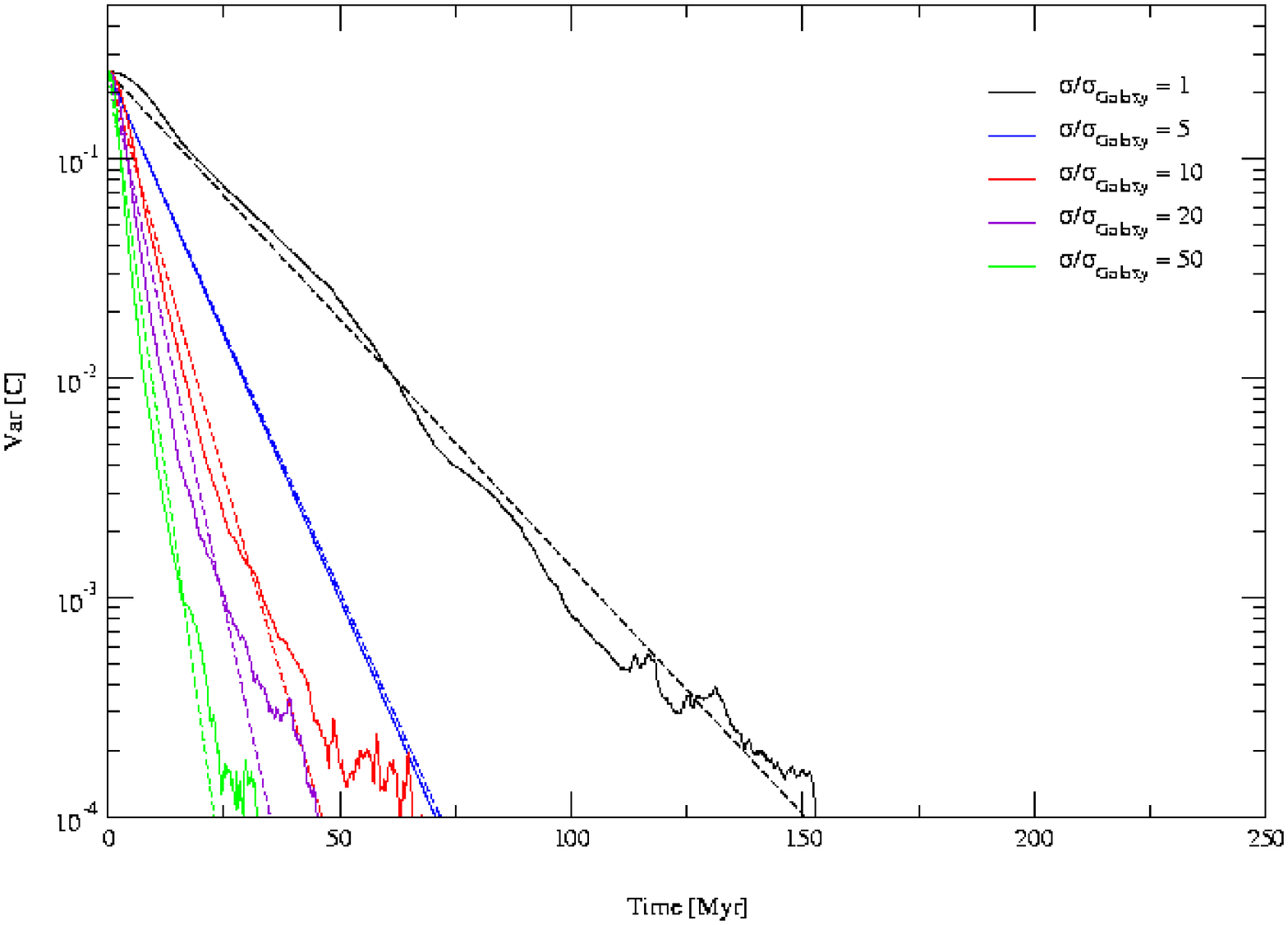}\\
\includegraphics[angle=-90,width=3in]{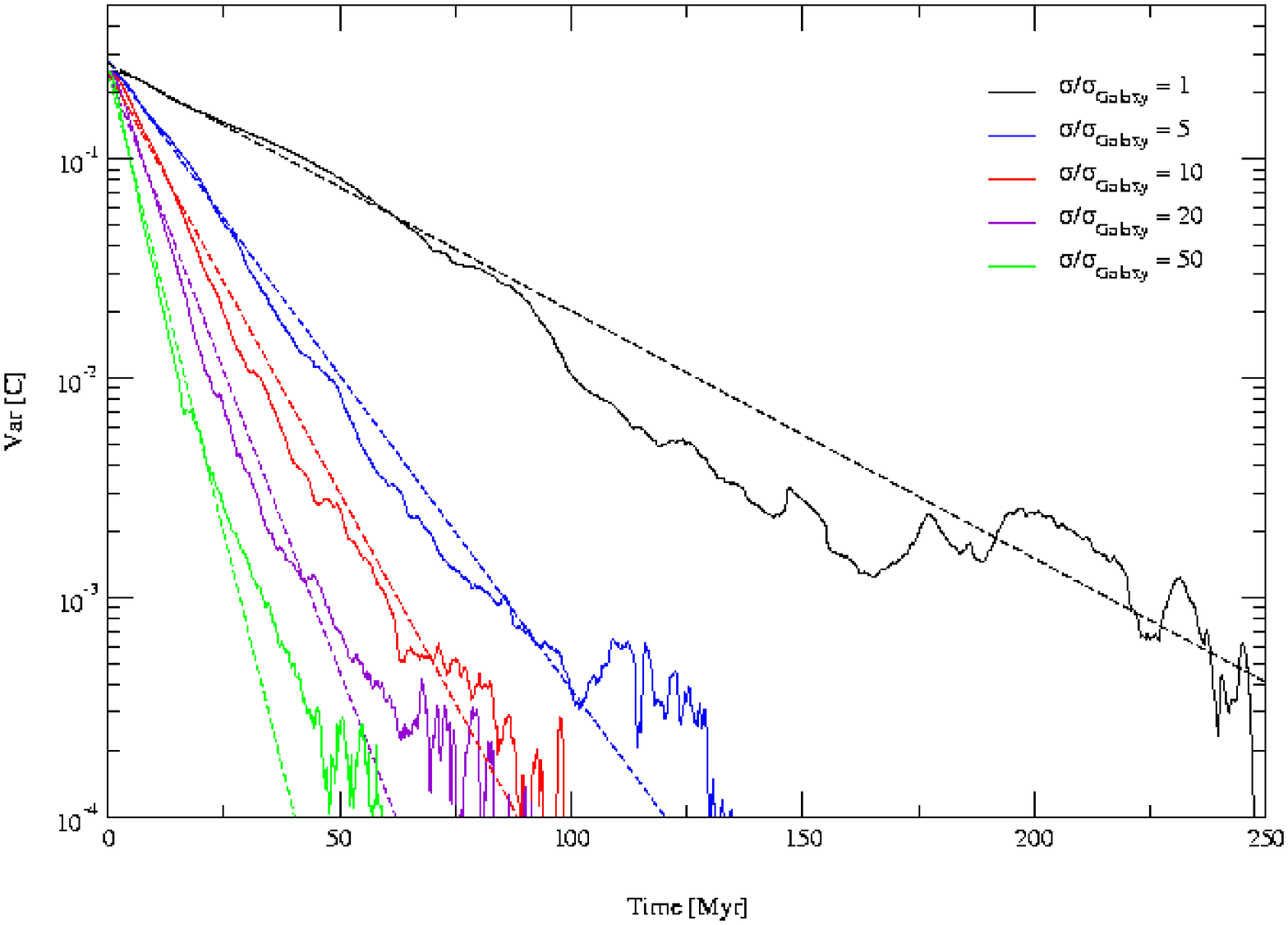}\\
\includegraphics[angle=-90,width=3in]{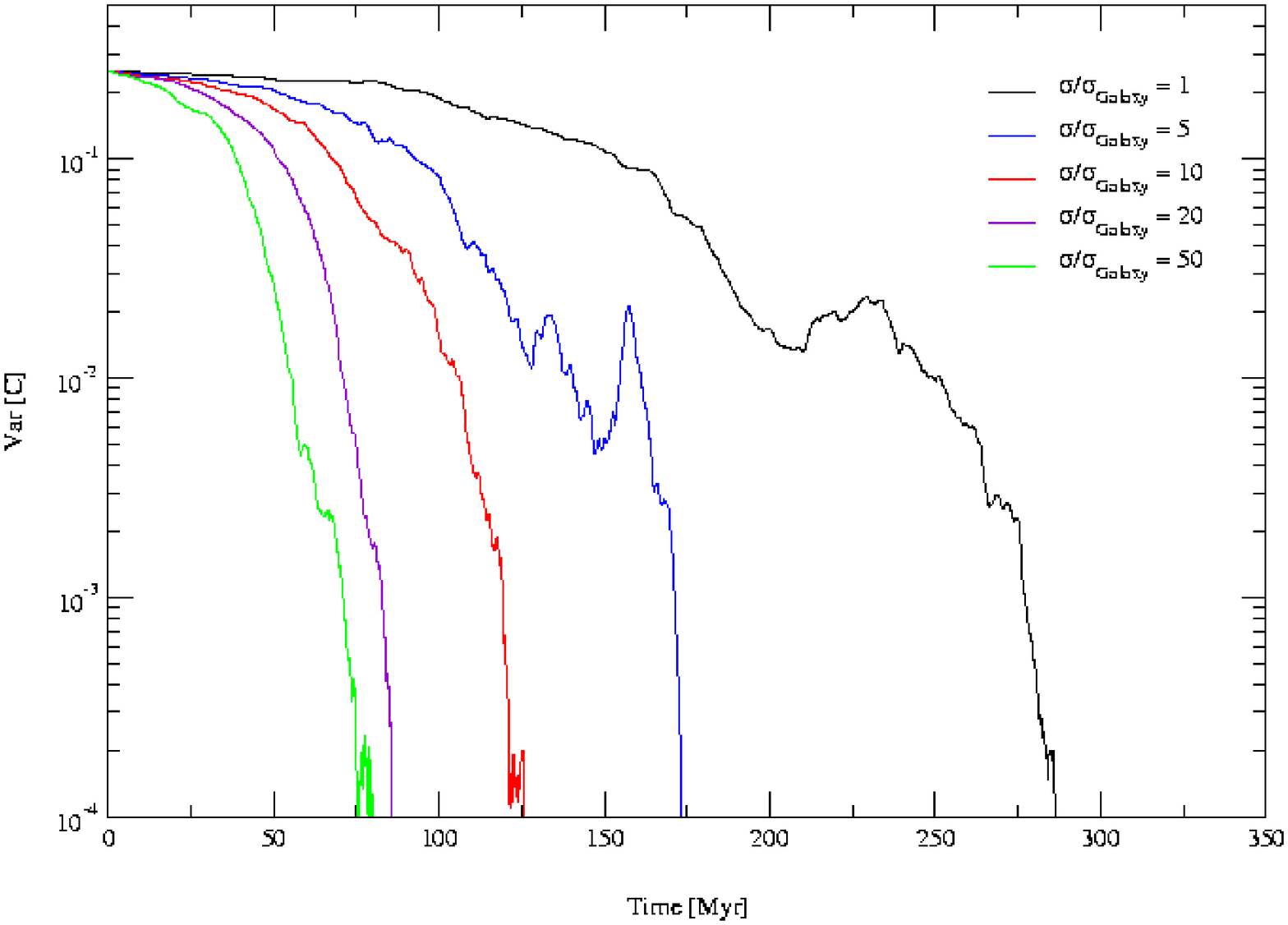}
\caption{Time evolution of the variance of C for inhomogeneities
having an initial length scale of 25 (top panel), 50 (middle panel)
and 500 pc (bottom panel). The plots show the variance (full lines)
and best fits to the variance (dashed lines) on the two first
panels. On the bottom panel we do not show any fit to the
variance. The fits are made down to $10^{-3}$ the smallest value at
which the variance decreased steadily.
}
\label{variance1}
\end{figure}

\begin{figure}
\centering
\includegraphics[angle=-90,width=3.3in]{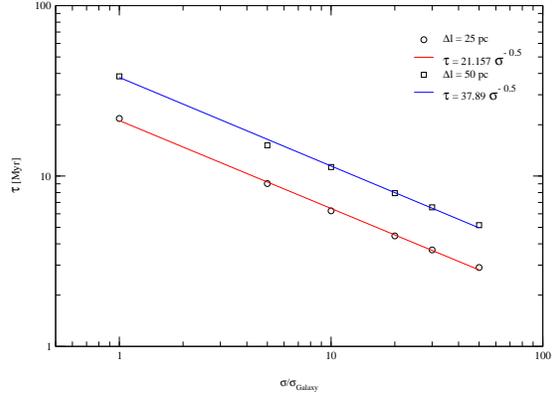}
\caption{Time variation of the time constant found in the fits for the variance of C for inhomogeneities $l=25$ and 50 pc.}
\label{variance2}
\end{figure}

\begin{figure}
\includegraphics[angle=-90,width=3.3in]{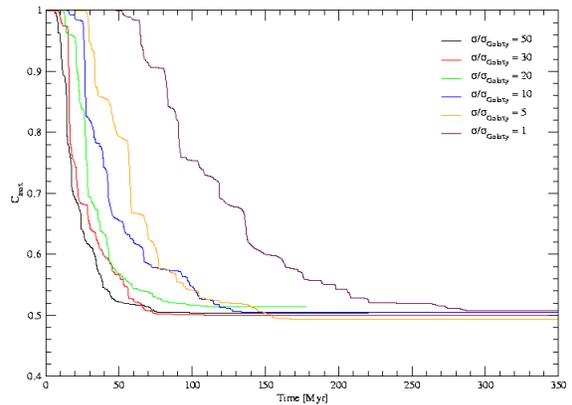}
\caption{Time evolution of the maximum value of the tracer field in a checkerboard, composed of 25 pc length squares, for $\sigma/\sigma_{\rm gal}=1$, 5, 10, 20, 30 and 50.
\label{cmaxtime}}
\end{figure}

\clearpage

\begin{figure}
\includegraphics[angle=-90,width=3.3in]{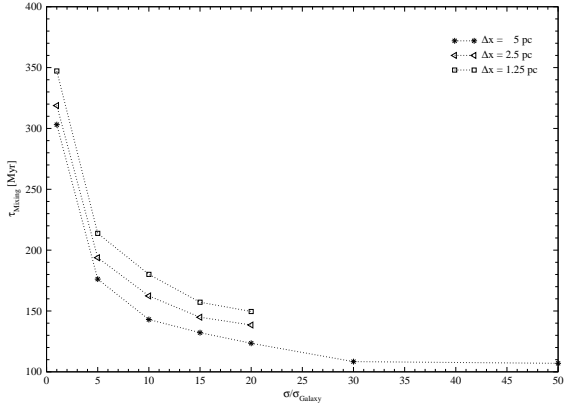}
\caption{Variation of mixing time of inhomogeneities with $l=50$ pc with SN rate $\sigma/\sigma_{\rm gal}$ for three finer grid resolutions $\Delta x=1.25,~2.5,~5$ pc.
\label{mixtimesnrate}}
\end{figure}

\begin{figure}
\includegraphics[angle=-90,width=3.3in]{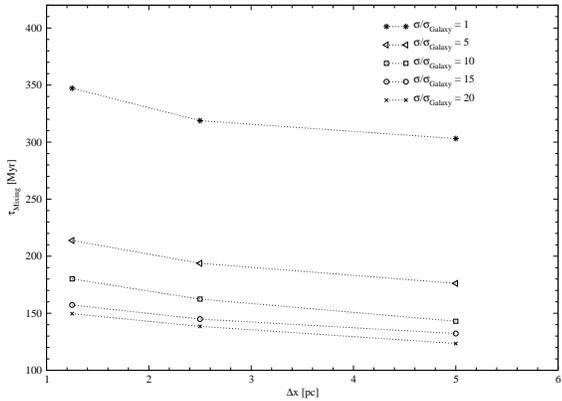}
\caption{Variation of mixing time of inhomogeneities with $l=50$ pc with grid resolution $\Delta x=1.25,~2.5,~5$ pc for $\sigma/\sigma_{\rm gal}=1,~5,~10,~15,~20$. There is a slight increase of the mixing time by factors between 1.052 to 1.2 when resolution is doubled.
\label{mixtimeresolution}}
\end{figure}

\begin{figure}
\centering
\includegraphics[angle=-90,width=3.4in]{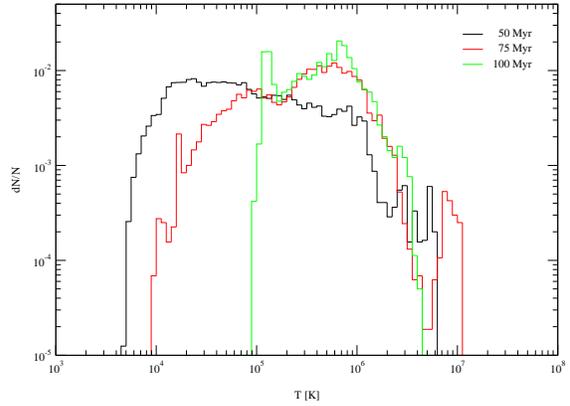}\\
\includegraphics[angle=-90,width=3.4in]{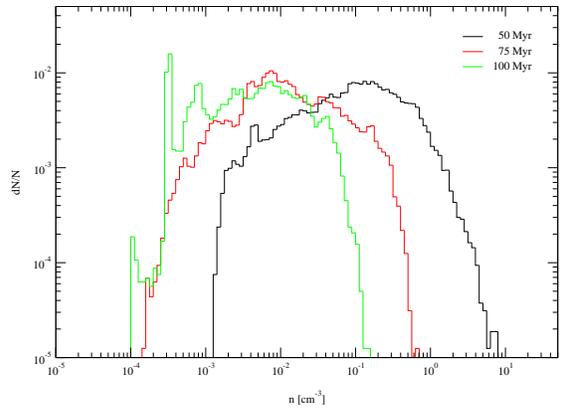}
\caption{Temperature and density PDFs of the disk gas taken at $-250\le z\le 250$ pc at times 50 (black), 75 (red) and 100 (green) Myr of disk evolution with a SN rate equal to 20 times that of the Galaxy. The PDFs show that the disk gas has a log-normal behaviour over a wide range of densities and temperatures. At 75 Myr the disk gas has a density smaller than 0.7 cm$^{-3}$, while at 100 Myr the disk gas has a density smaller than 0.1 cm$^{-3}$ and a temperature greater than $10^{5}$. 
\label{PDF-T}}
\end{figure}

\clearpage

\begin{figure*}
\centering
\includegraphics[angle=-90,width=2.8in]{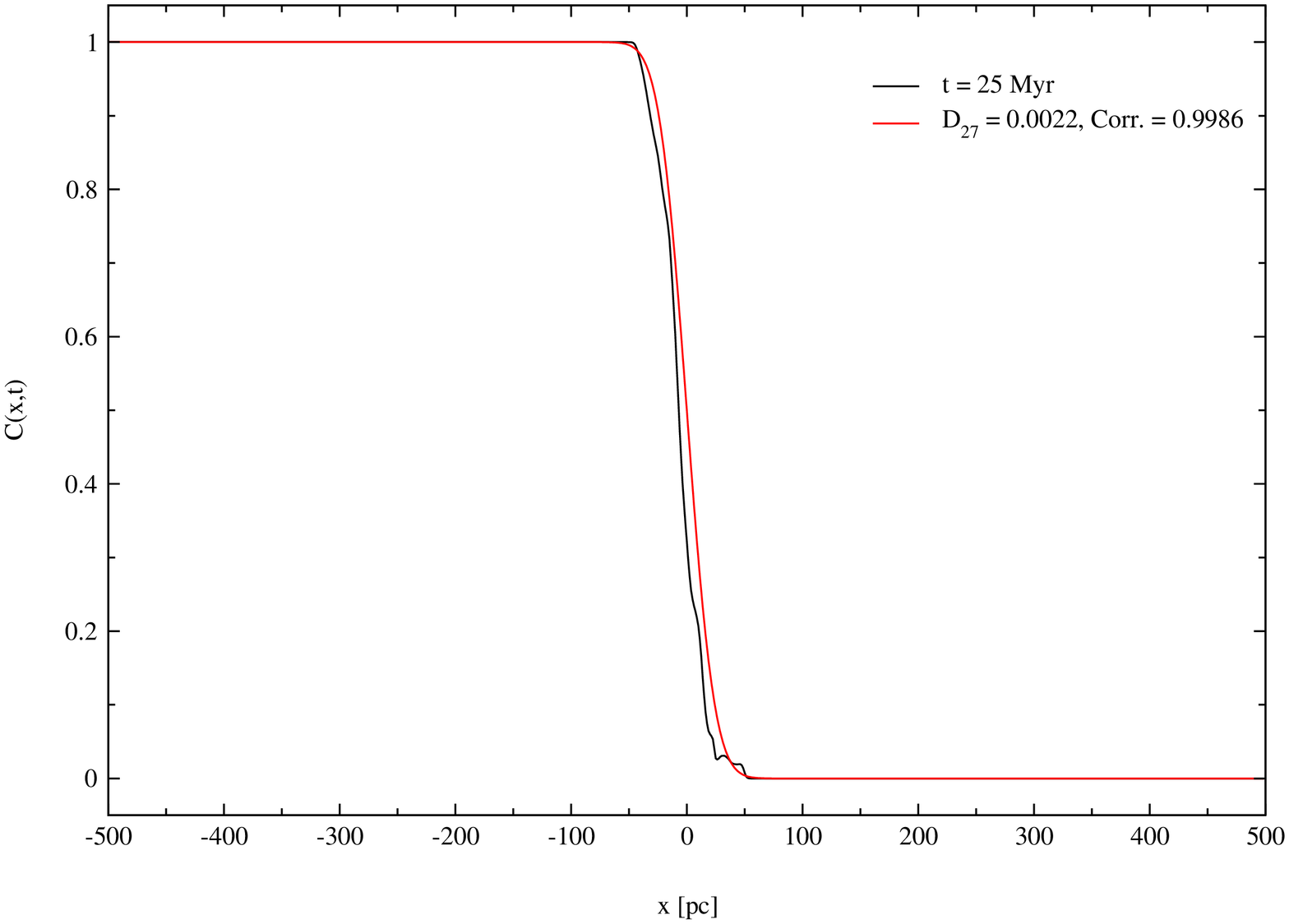}\includegraphics[angle=-90,width=2.8in]{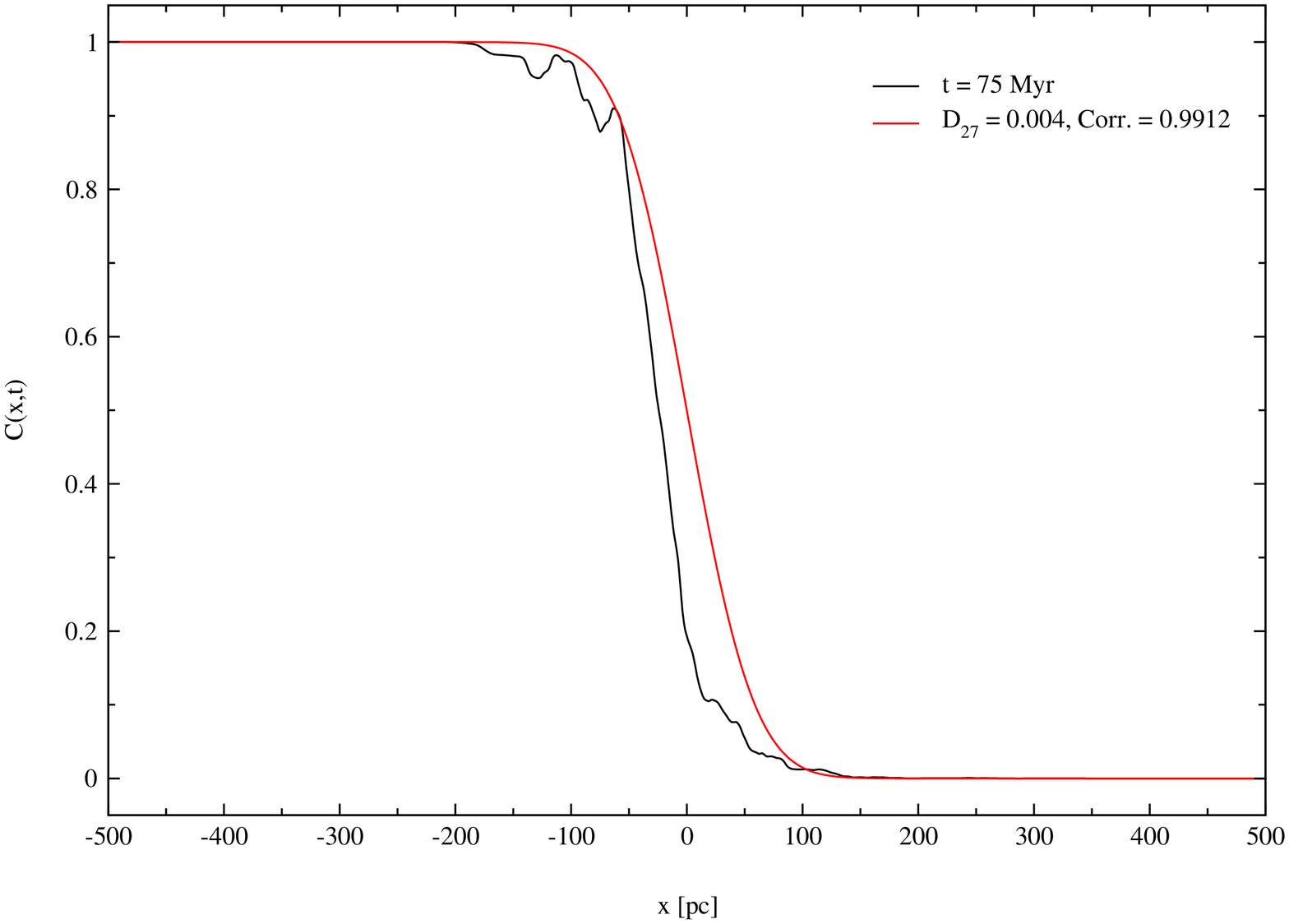}\\
\includegraphics[angle=-90,width=2.8in]{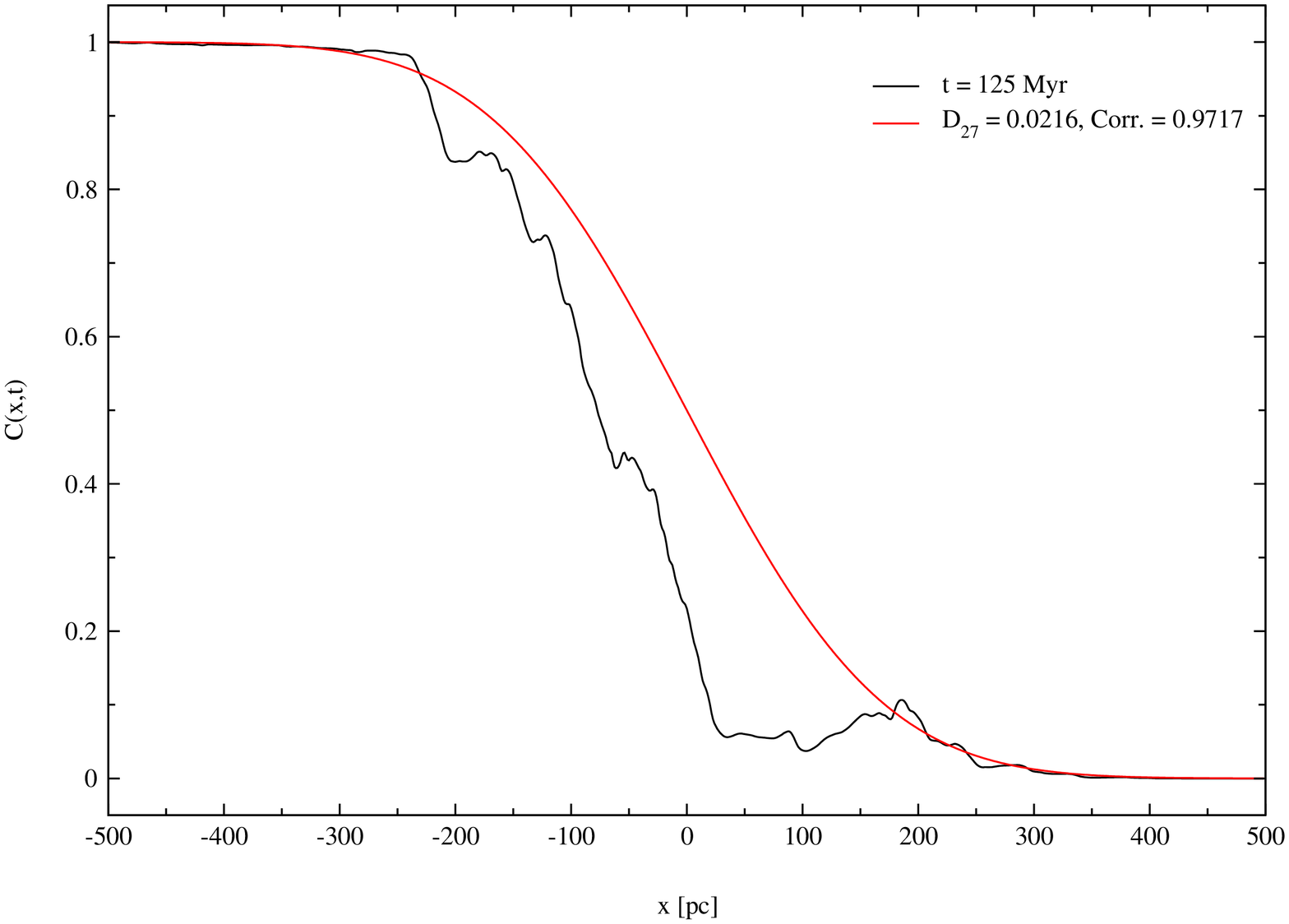}\includegraphics[angle=-90,width=2.8in]{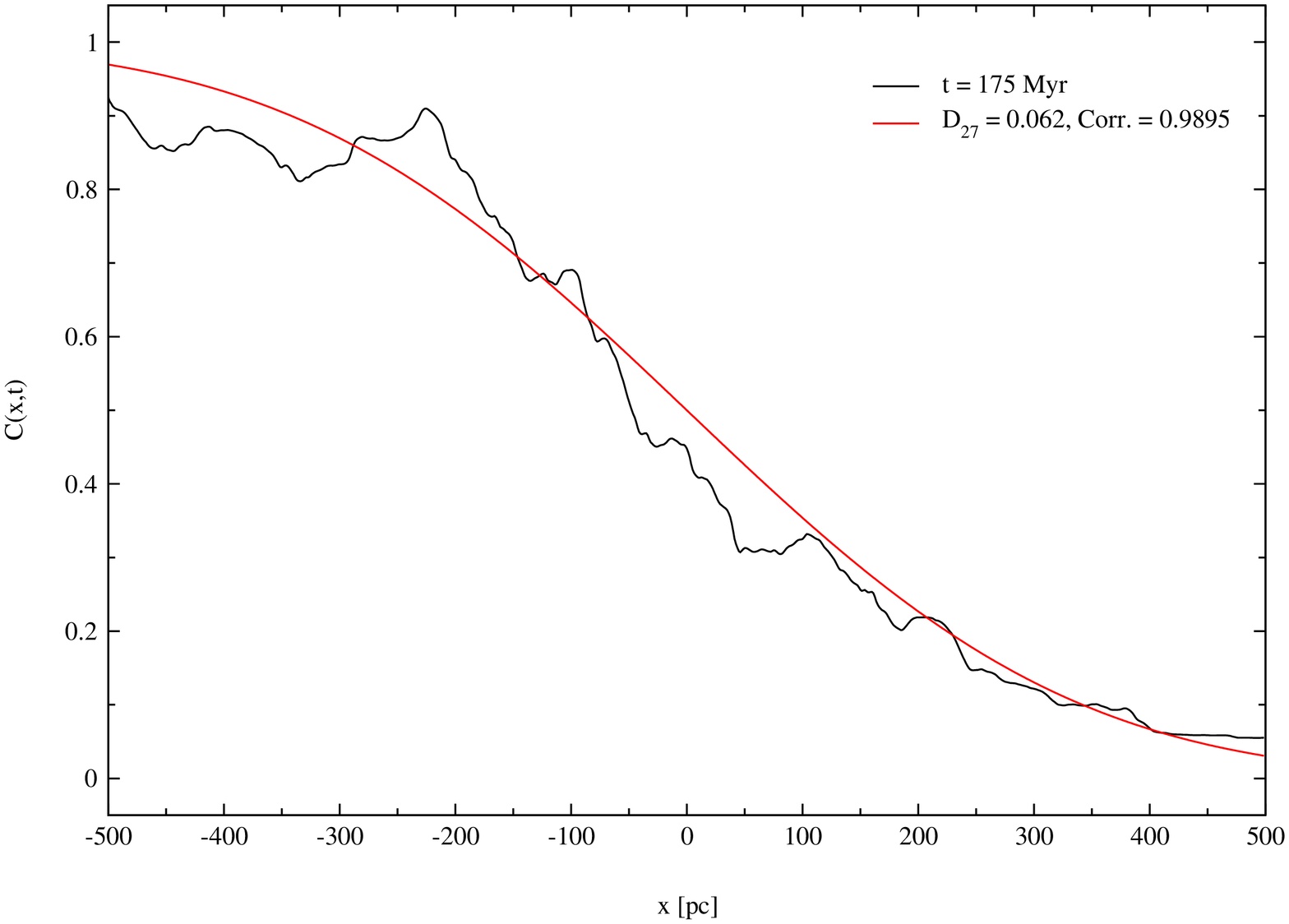}\\
\includegraphics[angle=-90,width=2.8in]{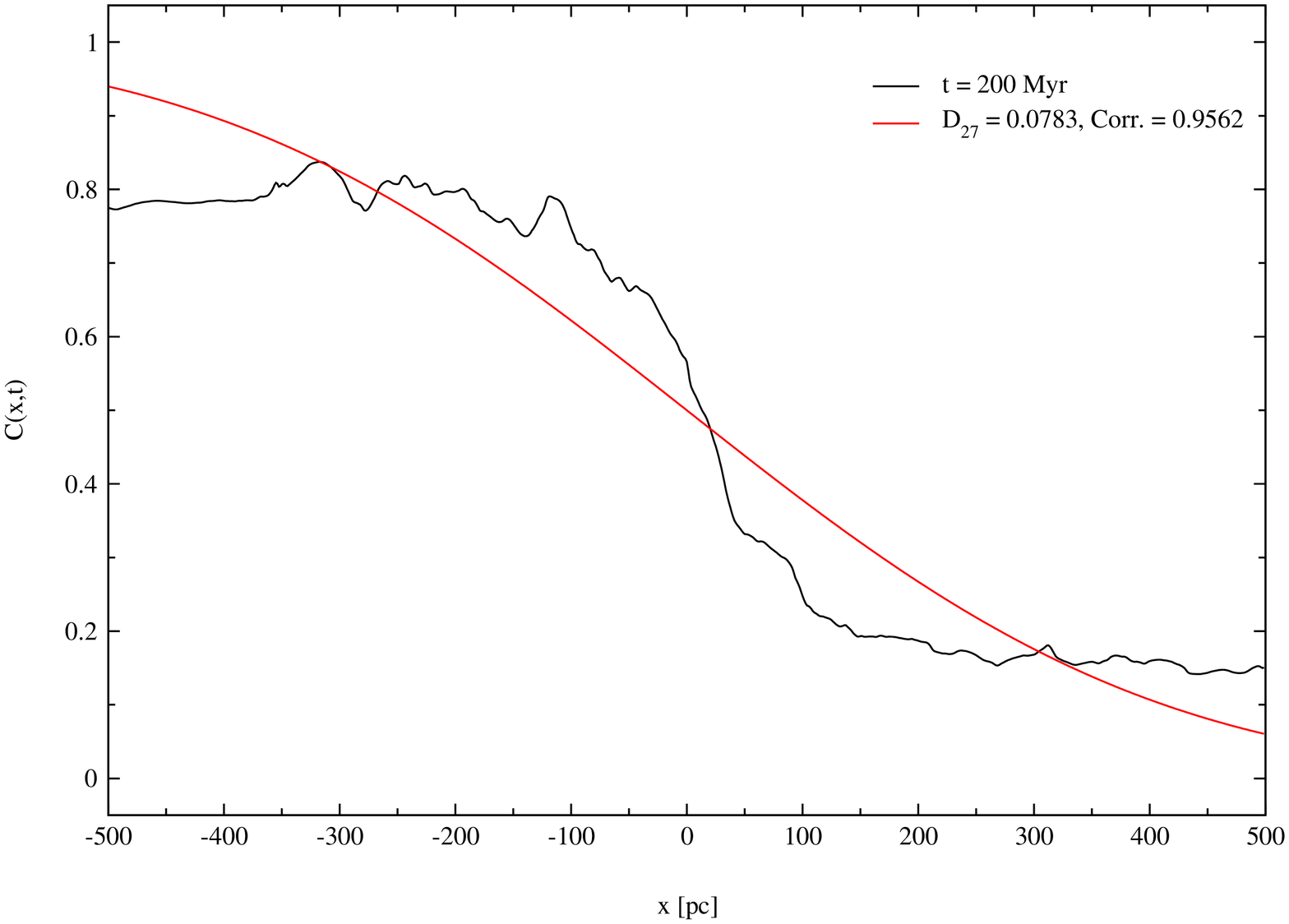}\includegraphics[angle=-90,width=2.8in]{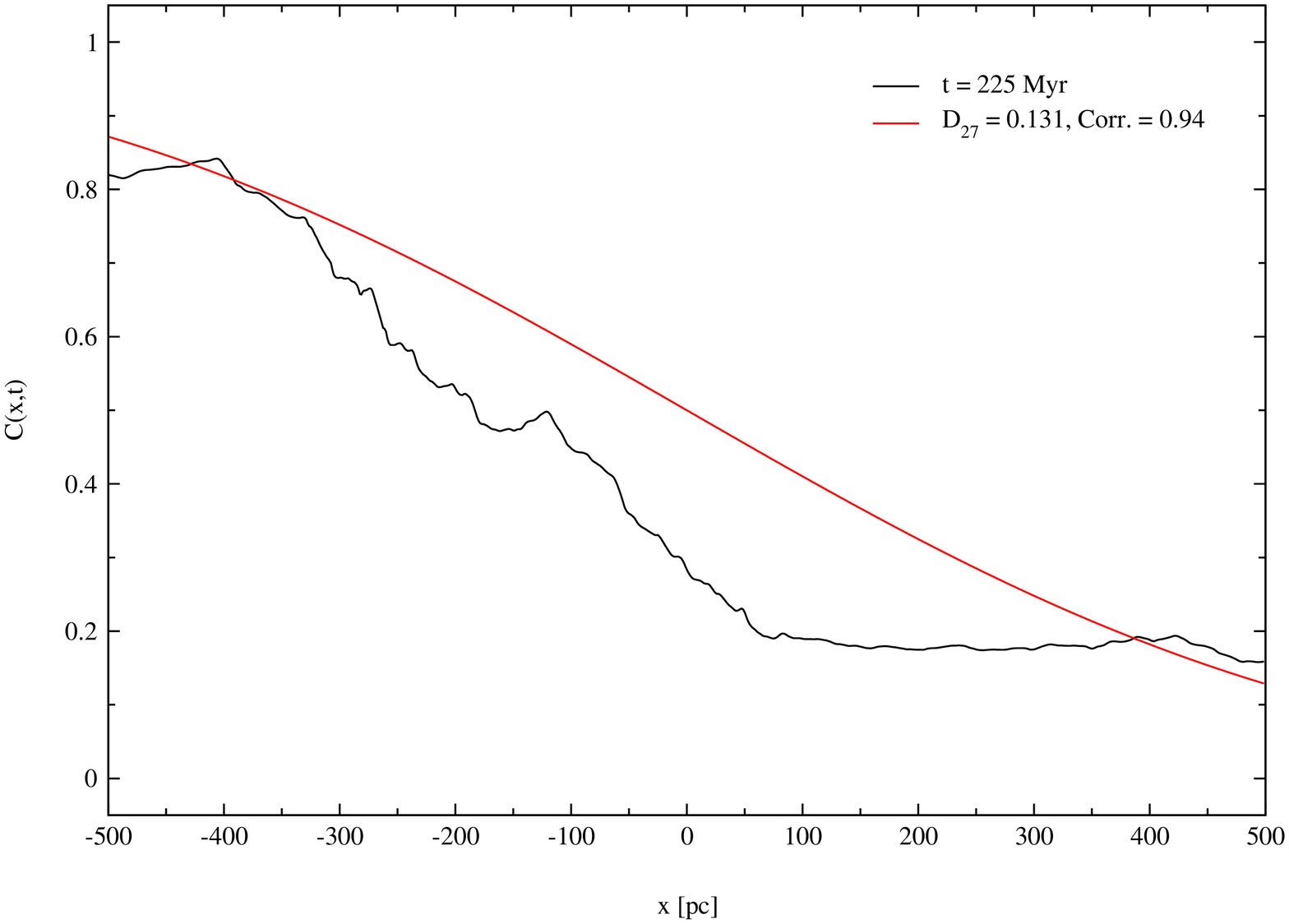}
\caption{Average value of the tracer field (black solid line),
measured for $y\geq 500$ and $0\leq x\leq 1000$ pc for the
checkerboard with squares having an initial length of 500 pc (shown in
the top panel of Figure 2), overlaid by a fit (red solid line) given
from equation~(\ref{solucao}) with $n=1$. The panels refer to 25, 75,
125, 175, 200 and 225 Myr of evolution of the tracer field in an ISM
with the Galactic SN rate. Note that, initially the fits follow the
average profile of the tracer field, but as time grows
equation~(\ref{solucao}) does not fit the profiles as well. This results from
the large number of small scale structures that are formed at the
interface region where the tracer field varies from 1 to 0. The
correlation factor varies from 0.998 at 25 Myr to 0.94 at 225 Myr.
\label{fitD1}}
\end{figure*}

\clearpage

\begin{figure}
\includegraphics[angle=-90,width=3.3in]{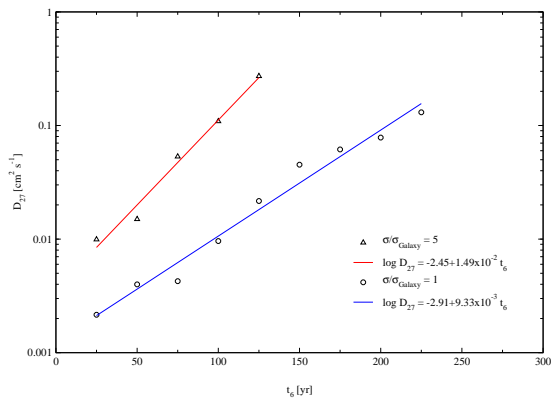}
\caption{Time variation of the diffusion coefficient D$_{27}$ for $\sigma/\sigma_{\rm gal}=1$ and 5. The straight lines are the best fit to the data points determined from the fits. The diffusion coefficient given in units of $10^{27}$ cm$^{2}$ s$^{-1}$, $D_{27}$, varies exponentially with time in units of Myr. 
\label{fitD}}
\end{figure}

\clearpage

\begin{table}
\caption{Summary of the simulation parameters}
\begin{tabular}{clrccc}
\hline
\hline
Run$^a$ & $\Delta x^{b}$ & $\sigma/\sigma_{\rm gal}^{c}$ &
$\tau_{25}^{d}$ & $\tau_{50}^{d}$ & $\tau_{500}^{d}$ \\ 
        &  [pc]  &      & [Myr] & [Myr] & [Myr]\\
\hline
L11  & 5    & 1  & ...  & 303.1 & 311.2 \\
L12  & 5    & 5  & ...  & 176.2 & 192.1 \\
L13  & 5    & 10 & ...  & 143.0 & 167.2 \\
L14  & 5    & 15 & ...  & 132.2 & 152.7 \\
L15  & 5    & 20 & ...  & 123.5 & 144.9 \\
L16  & 5    & 30 & ...  & 108.4 & 130.5 \\
L17  & 5    & 50 & ...  & 107.0 & 127.2 \\
L21  & 2.5  & 1  & 306.3 & 318.8 & 326.5 \\
L22  & 2.5  & 5  & 180.9 & 193.8 & 216.6 \\
L23  & 2.5  & 10 & 147.2 & 162.4 & 184.5 \\
L23  & 2.5  & 15 & 139.6 & 144.9 & 159.2 \\
L24  & 2.5  & 20 & 130.9 & 138.5 & 151.6 \\
L31  & 1.25 & 1  & ...  & 347.2 & 366.4 \\
L32  & 1.25 & 5  & ...  & 213.8 & 235.7 \\
L33  & 1.25 & 10 & ...  & 180.1 & 199.1 \\
L34  & 1.25 & 15 & ...  & 157.2 & 180.0 \\
L35  & 1.25 & 20 & ...  & 149.6 & 163.8 \\
\hline\\
\multicolumn{6}{l}{$^a$ Run label $Lab$; $a$ is the refinement level and $b$ the run number}\\
\multicolumn{6}{l}{$^b$ Resolution of the finest level of refinement}\\
\multicolumn{6}{l}{$^c$ SN rate in units of the SN Galactic rate}\\
\multicolumn{6}{l}{$^d$ Mixing time for inhomogeneities $l=25$, 50 and 500 pc}\\
\end{tabular}
\label{table1}
\end{table}
\clearpage

\begin{table}
\caption{Variation of the mixing time with resolution for inhomogenities with $l=50$ pc}
\begin{tabular}{cccccc}
\hline
\hline
$\sigma/\sigma_{\rm gal}^{a}$ & $\tau_{1.25}^{b}$ &  $\tau_{2.5}^{b}$ & $\tau_{5}^{b}$ & $\tau_{1.25}/\tau_{2.5}^{c}$ & $\tau_{2.5}/\tau_{5}^{c}$ \\
     &    [Myr]   & [Myr] & [Myr] &  & \\
\hline
 1   & 347.2  &  318.8 & 303.1 & 1.089 & 1.052 \\
 5   & 213.8  &  193.8 & 176.2 & 1.103 & 1.099 \\
 10  & 180.1  &  162.4 & 143.0 & 1.108 & 1.136 \\
 15  & 157.2  &  144.9 & 132.2 & 1.085 & 1.096 \\
 20  & 149.6  &  138.5 & 123.5 & 1.080 & 1.121 \\
\hline
\multicolumn{6}{l}{$^a$ SN rate in units of the SN Galactic rate}\\
\multicolumn{6}{l}{$^b$ Complete mixing time for 1.25, 2.5 and 5 pc}\\
\multicolumn{6}{l}{$^c$ Ratio between the mixing times for different resolutions}
\end{tabular}
\label{table2}
\end{table}

\end{document}